\documentclass[journal]{IEEEtran}  

\IEEEoverridecommandlockouts      

\usepackage{amsmath}
\allowdisplaybreaks[4]

\usepackage{graphics} 
\usepackage{epsfig} 
\usepackage{epic,eepic,epsfig}
\usepackage{epstopdf}
\usepackage{mathptmx} 
\usepackage{times} 
\usepackage{amsmath} 
\usepackage{amssymb}  
\usepackage{accents}
\usepackage{bm}
\usepackage{cancel}
\usepackage{mathrsfs}
\usepackage{nomencl}
\makenomenclature
\usepackage{longtable}
\usepackage{svg}

\usepackage{float}
\usepackage{pstricks}
\usepackage[normalem]{ulem}
\usepackage{hyperref}
\usepackage[capitalize, nameinlink]{cleveref}
\usepackage{algorithm}
\usepackage{algpseudocode}
\usepackage{multirow}
\usepackage{lipsum}
\usepackage{graphicx}
\usepackage{caption}
\usepackage{subcaption}
\usepackage{fancyhdr}   
\usepackage{soul}  

\DeclareMathOperator*{\argmin}{arg\,min}

\def\lungo #1{\mathord{\buildrel{\lower3pt\hbox{$\scriptscriptstyle\frown$}}
\over #1 } }

\def\1D#1#2{{{\partial}\over{\partial #2}}#1}

\def\1d#1#2{{{d}\over{d #2}}#1}

\newcommand{{\R}}{{\mathbb{R}}}
\newcommand{{\C}}{{\mathbb{C}}}

\newtheorem{remark}{Remark}

\begin{document}

\title{A Game-Theoretic Predictive Control Framework with Statistical Collision Avoidance Constraints for Autonomous Vehicle Overtaking}

\author{Sheng Yu, Boli Chen, Imad M. Jaimoukha, and Simos A. Evangelou
\thanks{S. Yu, I. M. Jaimoukha, and S. A. Evangelou are with the Department of Electrical and Electronic Engineering at Imperial College London, UK
        {\tt\small (sheng.yu17@imperial.ac.uk, i.jaimouka@imperial.ac.uk,\\ s.evangelou@imperial.ac.uk)}}%
\thanks{B. Chen is with the Department of Electronic and Electrical Engineering at University College London, UK
        {\tt\small (boli.chen@ucl.ac.uk)}}%
}

\thispagestyle{empty}
\setcounter{page}{0}
\begin{figure*}
\centering
\includegraphics[width=\textwidth]{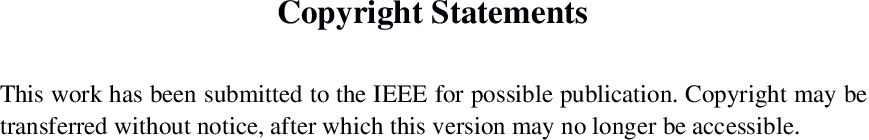}
\end{figure*}

\maketitle
\setlength{\headheight}{22.41992pt}
\thispagestyle{fancy}
\chead{This work has been submitted to the IEEE for possible publication. Copyright may be transferred without notice, after which this version may no longer be accessible.}
\rhead{~\thepage~}
\renewcommand{\headrulewidth}{0pt}

\pagestyle{fancy}
\chead{This work has been submitted to the IEEE for possible publication. Copyright may be transferred without notice, after which this version may no longer be accessible.
}
\rhead{~\thepage~}
\renewcommand{\headrulewidth}{0pt}

\maketitle
\begin{abstract}This work develops a control framework for the autonomous overtaking of connected and automated vehicles (CAVs) in a mixed traffic environment, where the overtaken vehicle is an unconnected but interactive human-driven vehicle. The proposed method, termed the Game-Theoretic, PRedictive Overtaking (GT-PRO) strategy, successfully decouples the longitudinal and lateral vehicle dynamics of the CAV and comprehensively coordinates these decoupled dynamics via innovative longitudinal and lateral model predictive (MPC) based controllers, respectively. To address the real-time interactive behavior of the human-driven overtaken vehicle, a dynamic Stackelberg game-based bilevel optimization is solved by the lateral controller to directly control the CAV lateral motion and predict the overtaken vehicle longitudinal responses that are subsequently shared with a stochastic MPC that governs the CAV longitudinal motion.
The proposed strategy exploits a comprehensive real-world dataset, which captures human driver responses when being overtaken, to tune the game-theoretic lateral controller according to the most common human responses, and to statistically characterize human uncertainties and hence implement a collision avoidance chance constraint for the stochastic longitudinal controller.
The simulation results for both polite and aggressive human response case studies of the overtaken vehicle demonstrate that the proposed GT-PRO can achieve for this range of human driver responsiveness, safer, more efficient, and more comfortable autonomous overtaking, as compared to existing autonomous overtaking approaches in the literature. Furthermore, the results suggest that the GT-PRO method is capable of real-time implementation.      
\end{abstract}


\section{INTRODUCTION}
\begin{figure}[t!]
\centering
\includegraphics[width=1\columnwidth]{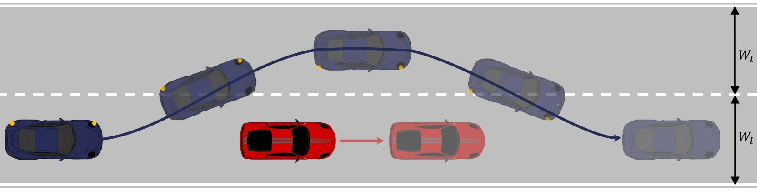}
\caption{Illustration of an overtaking scenario, where the dark blue vehicle represents the ego CAV, signaling to pull out into the overtaking lane. It then passes and merges back in front of the red overtaken vehicle. The uniform lane width is $W_{l}$.  
}
\label{fig:overtaking_intro}
\end{figure}

With the growing development in intelligent transportation systems (ITSs), connected and automated vehicles (CAVs) are becoming a notable part of the future ground traffic network, such as in cruising \cite{Yu2024TTE, Hamed2024}, intersection \cite{PanXiao2023}, parking \cite{Lian2023}, and overtaking \cite{Gao2022} scenarios, with benefits on traffic passing efficiency, ecological driving, and safety \cite{Guanetti2018, Rahman2023, Ortega2024}.   
Among these, the overtaking maneuver, typically composed of pulling-out, passing, and cutting-in phases, as illustrated in Fig.~\ref{fig:overtaking_intro}, stands out as a frequent and safety-critical challenge in autonomous driving. Advances in control and optimization techniques for autonomous overtaking have also been extended to tackle more complex driving scenarios \cite{Ortega2024, Lodhi2023}.

Among various control methods to achieve safe and comfortable autonomous overtaking maneuvers in the literature, model predictive control (MPC) is widely adopted due to its advantages in handling constraints \cite{Dixit2018}, such as a data-driven MPC \cite{Liyiqun2024}, a robust MPC \cite{Dixit2020}, a control barrier function-based MPC \cite{Yuan2024}, and a learning-based MPC \cite{Yuan2025}. 
Although MPC is widely applied in overtaking scenarios, including efforts on linearisation, vehicle dynamics modeling, and collision avoidance, modeling the interaction between the overtaking CAV and the vehicle being overtaken remains a key challenge in establishing appropriate collision avoidance constraints, due to the dynamic and interactive nature of the overtaking maneuver. 
A common approach in the literature is to assume the velocities of the overtaken vehicles are either available to the CAV or constant \cite{Gratzer2024,Karlsson2020}. Ref \cite{Gratzer2024} relies on vehicle-to-everything (V2X) communication, where the future motions of overtaken vehicles are available. It approximates the vehicle occupancy as two circles centered at the midpoints of the front and rear axles and implements two collision avoidance constraints directly to these two unit circles, respectively. The overtaking framework proposed in \cite{Karlsson2020} adopts an assumption that the longitudinal speed of the leading vehicle is known and constant. It then introduces speed-independent rectangular regions to model an overtaking window that is safe for the CAV and a critical area around the overtaken vehicle, which defines the feasible region of the overtaking trajectory. In addition to traditional rectangles and circles, an elliptical constraint is adopted by \cite{Wang2024} to keep a safe distance between the ego vehicle and nearby obstacles, where the parameters of the ellipse are closely related to the dimensions of the surrounding vehicles. Moreover, \cite{Chai2023} approximates each vehicle occupied area as a convex polygon, and to ensure the avoidance of any crash, it is expected that all corner points of one polygon are located outside of the polygon associated with the other vehicle, such that the occupied areas of the two vehicles do not overlap. These approaches eliminate uncertainties related to overtaken vehicle velocities in collision avoidance constraint design, by presuming complete knowledge of those velocities. Although this assumption is reasonable in fully autonomous traffic where all vehicles are CAVs with information exchange based on the V2X network, it is expected that 
a mixed traffic with CAVs coexisting with human-driven vehicles (HDVs) on the road will remain the dominant traffic mode for some time, as the autonomous driving technology and infrastructure mature and are deployed \cite{Chen2020}. As such, one of the significant characteristics of mixed traffic environments is the lack of accurate surrounding HDV information, including their position, velocity, and acceleration profiles. To make matters worse, the above methods overlook the impact of CAV control on the behavior of the HDV, which can lead to suboptimal and unsafe planning and control for CAVs.
In particular, in contrast to the cooperation that occurs between CAVs, the responses of HDVs in interactions between CAVs and HDVs 
are uncontrolled and can be non-cooperative. 
Such responses are difficult to predict with high confidence in a real-time manner.

Attempts to predict the interactive responses of HDV have been made in the literature by game-theoretic methods, which are introduced in ITS applications to handle uncertain surrounding vehicle velocities, such as in turning \cite{Rahmati2022}, merging \cite{Wei2022}, and overtaking \cite{Gao2022, Lizhuoren2024} maneuvers, by mimicking interactions between CAVs and HDVs. Specifically, \cite{Gao2022} presents a risk-aware reachability analysis based on martingale theory, such that the method proposed in \cite{Gao2022} is capable of overtaking an obstacle vehicle with uncertain motion. The collision avoidance constraint adopted in \cite{Gao2022} is based on an exact reformulation method that converts nondifferentiable boundaries into smooth and differentiable constraints as originated in \cite{Zhang2021}. Nevertheless, the original optimization problem faces feasibility issues as it balances the feasibility of automated overtaking against the risk of collision throughout the prediction horizon. The optimization problem is then reformulated into a solvable two-stage optimization problem addressing the two boundary cases, resulting in either a conservative or aggressive solution. In \cite{Lizhuoren2024}, an interaction-based decision-making process is used to model uncertain behaviors and motions of overtaken vehicles with their probabilistic action policy, where the uncertainties of HDV trajectories are described by a multivariate Gaussian distribution. Moreover, in \cite{Wei2022}, a Stackelberg game-based decision block generates optimal discrete strategies, followed by an MPC motion planning block that further determines control sequences. In addition, \cite{Zhang2024} proposes a dynamic Stackelberg game approach to control vehicle motions in lane-changing maneuvers.
In a Stackelberg game, the leader makes the initial move, knowing how the follower would react. After observing the leader's action, the follower responds accordingly \cite{Stackelberg2011}. As the overtaking CAV is the sole controlled entity and the obstacle HDV operates uncontrolled, the overtaking dynamic aligns naturally with the Stackelberg framework. By positioning the CAV as the leader, one can strategically control its actions to influence the HDV, the follower, behavior in a beneficial way, ensuring safer and more efficient overtaking maneuvers. Nevertheless, the combination of the Stackelberg game and dynamic optimization problems, such as MPCs, usually results in constrained bilevel optimizations, which are difficult to solve with real-time computational efficiency. While \cite{Wei2022} uses exhaustive search to solve the resulting bilevel optimization of the merging scenario because its control variables are discrete decision-making strategies, for problems seeking optimal control sequences, system dynamics are usually simplified for the convenience of computation. For example, in \cite{Zhang2024}, lateral movement control is simplified to making instantaneous lane change decisions between two lanes, by disregarding the vehicle lateral dynamics.
In addition, traditional game-theoretic approaches in the literature typically assume that the follower's best response is accurately known without considering any deviation between the game-predicted and real follower responses.

A robust control solution for autonomous overtaking that integrates Stackelberg game theory and dynamic motion control, without having to sacrifice significant fidelity of the system dynamics, does not currently exists. 
With the aim of addressing this identified research gap, this work tackles the safety-guaranteed autonomous overtaking problem involving complex interactions between the ego CAV and a surrounding HDV, as shown in Fig.~\ref{fig:overtaking_intro}. A novel control framework, namely Game-Theoretic, PRedictive Overtaking (GT-PRO), is proposed, bringing together game-theoretical with stochastic MPC techniques, and incorporating statistical results obtained from real-world human driver data for accurately capturing human behavioral uncertainties.
Specifically, this paper makes the following contributions:
\begin{itemize} 
    \item Compared with the authors' preliminary work \cite{Yu2024ECC} that merely controls the lateral motion of the CAV, while assuming its longitudinal speed is constant, in the present work, full CAV dynamics of appropriate fidelity are decoupled into lateral and longitudinal linearized dynamics and fully controlled by innovative lateral and longitudinal MPC-based controllers, respectively.
    A novel adaptive geometric collision avoidance constraint between the CAV and HDV, depending on the longitudinal velocities of both vehicles in contrast to \cite{Yu2024ECC} in which it depends only on the HDV longitudinal velocity, is also defined in a dynamic coordinate system, which convexifies the feasible region of the CAV's trajectory during the overtaking maneuver. 
    \item A dynamic Stackelberg game-based MPC, resulting in a constrained bilevel optimization, is formulated and solved with real-time computation efficiency for the lateral controller, generating optimal control sequences of the CAV lateral motion and predicting associated best responses of the HDV longitudinal motion in overtaking. In contrast to adopting and modeling a specific human longitudinal driving behavior that appears reasonable in a generic sense as in \cite{Yu2024ECC}, the best response modeling of the HDV follower in the GT-PRO is tuned to provide more realistic responses by aligning with the most common human driver decisions in overtaking according to real-world data of naturalistic vehicle trajectories recorded on German highways \cite{highDdataset}.
    \item A stochastic MPC is designed for the longitudinal controller to account for deviations between the game-theoretical and real HDV behaviors, which are caused by unbounded human uncertainties. The human uncertainties are statistically modeled based on the variance of human longitudinal driving behavior manifested in the real-world dataset \cite{highDdataset}, and according to this modeling, the collision avoidance constraint is converted into a chance constraint and integrated into the stochastic MPC. 
    \item Rather than comparing with empirical control rule-based solutions as found in many existing works \cite{Kesting2007}, comprehensive simulations are conducted to evaluate the safety, traffic passing efficiency, and comfort of GT-PRO against several state-of-the-art control methods under both polite and aggressive HDV behaviors. 
\end{itemize}

The rest of the paper begins with a statement of the autonomous overtaking problem in Section~\ref{sec:system description}, which also introduces the vehicle bicycle model and the proposed collision avoidance constraint setup. In Section~\ref{sec:control design}, the proposed overtaking control framework is described by introducing the game-theoretic MPC for the EV lateral dynamics, followed by the statistical-based stochastic MPC for the EV longitudinal dynamics. Simulation results of the proposed overtaking control framework and benchmark comparisons are illustrated and discussed in Section~\ref{sec:simulation_results}. Finally, conclusions are provided and a future work plan is suggested in Section~\ref{sec:conclusions}.

\emph{Notation}: Let $\mathbb{R}$, $\mathbb{R}_{>0}$ and $\mathbb{N}$ denote the real, the strictly positive real, and natural sets of numbers, respectively. For any two integers $m$ and $n$ satisfying $m\leq n$, $\mathbb{N}_{[m,n]}=\{m,m+1, \dots,n\}$.
$\mathbb{R}^n$, $\mathbb{R}^{n\times m}$, and $\mathbb{D}^{n}$ denote the space of an $n$-dimensional real (column) vector, an $n \times m$ real matrix, and an $n \times n$ diagonal matrix, respectively.
$\mathbb{S}^{n}_{\succ}$ and $\mathbb{S}^{n}_{\succeq}$ are sets of symmetric positive definite and positive semi-definite $n \times n$ matrices, 
respectively. 
$\mathbf{0}^{n\times m}$ is an $n \times m$ matrix with all zeros.
$A^{\top}$ denotes the transpose of $A$. 
\section{Problem Statement}
\label{sec:system description}

This article focuses on an overtaking scenario occurring on a two-lane, one-way straight road with uniform lane width $W_l$, as illustrated in Fig. \ref{fig:overtaking_intro}. The ego vehicle (EV, CAV) initiates the maneuver in its original lane, shifts into the overtaking lane to pass a slow overtaken vehicle (OV, HDV), and then merges back into the initial lane. The objective is for the EV to complete the overtaking maneuver safely, efficiently (in terms of time in the overtaking lane), and comfortably. Without loss of generality, this study considers right-hand traffic, where the overtaking lane is on the left side of the initial lane.

\subsection{Vehicle Kinematic Model}
\label{sec:vehicle_kinematic_model}
\begin{figure}[h!]
\centering
\includegraphics[width=1\columnwidth]{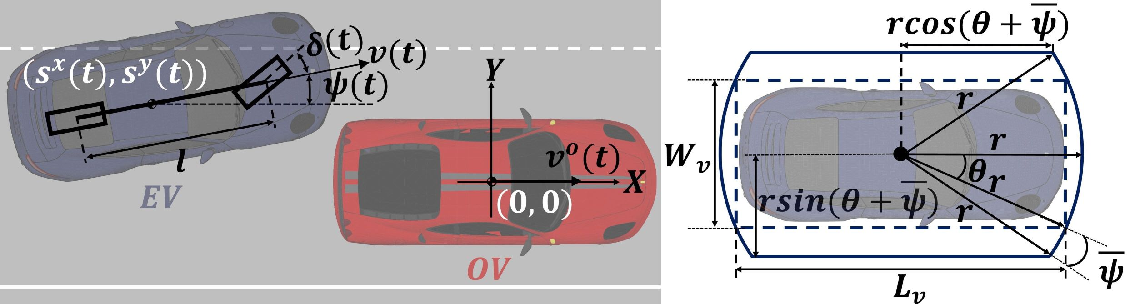}
\caption{Illustrations of the kinematic bicycle model and the moving coordinate system utilized to model vehicle motions (left), as well as the arc-polygon-shape vehicle space occupation adopted to set up collision avoidance constraints (right).}
\label{fig:bicycle_model_arcpolyton}
\end{figure}
\begin{figure*}[h!]
\centering
\includegraphics[width=\textwidth]{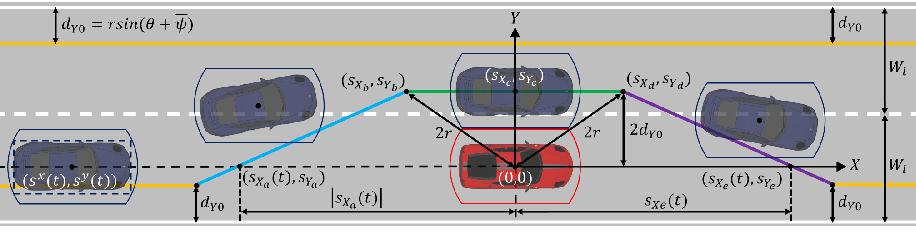}
\caption{A geometric illustration of the proposed collision avoidance constraint setup. The red OV is centered at the origin (0,0) of a moving $X$-$Y$ coordinate system. 
The dark blue vehicles illustrate possible EV occupancies during the overtaking, with the current position of the EV center shown at point ($s^{x}(t),s^{y}(t)$). 
Dark blue and red arc-polygons represent possible EV and OV occupancies, respectively.
Road boundaries and road lane markings are denoted by solid and dashed white lines, respectively. 
Blue, green, and purple segments represent the EV center collision avoidance boundaries for the various overtaking phases, which are explicitly formulated in \eqref{eq:col_avod} and \eqref{eq:col_avod_numerical_expressions}.
In addition, yellow segments are EV center constraints associated with the lane boundaries, as formulated in \eqref{eq:col_avod_roadside}. 
}
\label{fig:constraint_dynamic}
\end{figure*}

This work utilizes a one-track kinematic bicycle model to simulate the motions of the EV during overtaking, which keeps a balance between modeling accuracy and control simplicity \cite{Rajamani2012}. 
Moreover, for ease of modeling vehicle interactions, a moving coordinate system with its origin $(0,0)$ fixed to the geometric center of the OV is adopted in this work, as illustrated in Fig. \ref{fig:bicycle_model_arcpolyton}. The $X$- and $Y$-axes denote the longitudinal and lateral directions of the OV, respectively. The OV (and its coordinate system) moves forward by a speed of $v^{o}(t)$ according to the discrete time dynamic
\begin{equation}
    \begin{aligned}
        v^{o}(t+1)&=v^{o}(t)+a^{o}(t)\Delta T, \label{eq:vehicle_bicycle_model5}
    \end{aligned}
\end{equation}
where $a^{o}(t)$ represents the total acceleration (positive) or deceleration (negative) 
of the OV. $\Delta T \in \mathbb{R}_{>0}$ is the sampling interval and $t\in \mathbb{N}_{[1,N_T]}$ is the sampling index, with $N_T=T/\Delta T$ the total number of samples and $T$ the predefined total simulation time. The OV is assumed to maintain a straight trajectory along its longitudinal direction, shown in Fig. \ref{fig:bicycle_model_arcpolyton}, while being overtaken, hence, its lateral motion is not considered. The coordinates of the geometric center of the EV in the moving coordinate system attached to the OV are $s^{x}(t)$ and $s^{y}(t)$, and the EV heading angle with respect to the OV longitudinal direction is $\psi(t)$, as shown in Fig. \ref{fig:bicycle_model_arcpolyton}. The EV side slip angle is assumed to be neglected such that the EV geometric center forward velocity, $v(t)$, is along the EV's longitudinal axis. 
The equations of motion of the EV are
\begin{subequations}\label{eq:vehicle_bicycle_models}
    \begin{align}
         s^{x}(t+1)&=s^{x}(t)+\left( v(t) \cos(\psi(t))-v^{o}(t)\right)\Delta T, \label{eq:vehicle_bicycle_model1}\\
         s^{y}(t+1)&=s^{y}(t)+v(t) \sin (\psi(t))  \Delta T,\label{eq:vehicle_bicycle_model2}\\
         \psi(t+1)&=\psi(t)+\frac{v(t) \tan(\delta(t))}{l} \Delta T, \label{eq:vehicle_bicycle_model3}\\
         {v}(t+1)&=\!v(t)\!+a(t)\Delta T,  \label{eq:vehicle_bicycle_model4}
    \end{align}
\end{subequations}
where $l$ is the EV wheelbase. $a(t)$ and $\delta(t)$, which are the control inputs in the overtaking problem, are the EV geometric center forward acceleration and front wheel steering angle, respectively.   
In this work, \eqref{eq:vehicle_bicycle_model5} and \eqref{eq:vehicle_bicycle_models} are directly employed for numerical simulation of the vehicle motion later in Section \ref{sec:simulation_results}, and they are further used to design the control scheme after adopting small-angle and decoupling of longitudinal and lateral dynamics approximations, which will be further explained in Section \ref{sec:control design}.  
 
\subsection{Overtaking Collision Avoidance Constraints}
\label{sec:collision_avoidance_constraints}
The proposed collision avoidance constraint formulation begins by representing the occupancy of the vehicle. As shown by the dashed box in Fig. \ref{fig:bicycle_model_arcpolyton}, the vehicle occupancy can be traditionally approximated as a horizontal rectangle of length $L_v$ and width $W_v$, where $2r=\sqrt{L_{v}^{2}+W_{v}^{2}}$ is the rectangle diagonal length and $\theta=\arctan(\frac{W_{v}}{L_{v}})$ denotes the angle between the diagonal and the long side of the rectangle.

To account for the additional space occupied when the vehicle steers, the occupancy is expanded from the horizontal rectangle to a horizontal arc-polygon enclosed by solid straight line segments and arcs with a radius of $r$, as shown in Fig. \ref{fig:bicycle_model_arcpolyton}. 
The angle of the arc is determined by the vehicle characteristic ($\theta$) and the maximum heading angle ($\overline{\psi}$). 
The length and the width of the arc-polygon are $2r$ and $2d_{Y_{0}}$, respectively, with $d_{Y_{0}}=r\sin{(\theta+\overline{\psi})}$.
The side of the arc-polygon is kept parallel to the road even when the vehicle steers, and collisions can be avoided as long as the arc-polygons of the two vehicles do not overlap. This approach is much less conservative compared to the circle-based representation \cite{Gratzer2024}, which is commonly used in wheeled robots. In contrast to the conventional rectangle method \cite{Gao2022}, which employs a nonlinear rotation matrix to account for additional occupancy during vehicle steering, the proposed arc-polygon approach simplifies the calculation process and provides a basis for developing linear collision avoidance constraints, as will now be explained. 

Based on the arc-polygon definition, to satisfy the collision avoidance criteria, the geometric center point of the EV should remain within the area bounded by the blue, green, purple, and yellow segments, shown in Fig. \ref{fig:constraint_dynamic}, for the whole overtaking maneuver. The blue, green, and purple segments are related to collision avoidance with the OV. 
The line segments are given by the points ($s_{X_{n}}, s_{Y_{n}}$) $\forall n \in \{a, b, c, d, e\}$, respectively, which are determined as follows in the moving coordinate system:
\paragraph{Before and after overtaking (blue and purple segments)} The blue segment that passes through points ($s_{X_{a}}(t), s_{Y_{a}}$) and ($s_{X_{b}}, s_{Y_{b}}$) is implemented to avoid rear-end collisions when the EV changes from the original lane to the overtaking lane before passing the OV. Similarly, the purple segment that passes through points ($s_{X_{d}}, s_{Y_{d}}$) and ($s_{X_{e}}(t), s_{Y_{e}}$) guarantees the avoidance of rear-end collisions when the EV merges back to the original lane after passing the OV. The 
coordinates 
\begin{subequations}\label{eq:coordinates}
    \begin{align}
        s_{X_{a}}(t)&=-\left(d_{X_{0}}+v(t)\underline{t}\right), \label{eq:coordinates_1}\\
        s_{X_{e}}(t)&=d_{X_{0}}+v^{o}(t)\underline{t}, \label{eq:coordinates_2}
    \end{align}
\end{subequations}
are defined to contain a constant part, such that the collision avoidance constraints 
enable a minimum longitudinal distance of $d_{X_0}$ between the geometric centers of the two vehicles at standstill, and a time-varying part dependent on the 
EV and OV velocities, $v(t)$ and $v^{o}(t)$, respectively, such that 
there is a minimum car-following headway time, $\underline{t}$, to react effectively to the front vehicle braking. 
\paragraph{During overtaking (green segment)} The horizontal green segment passing through the point ($s_{X_{c}}, s_{Y_{c}}$) ensures the avoidance of side collisions when the EV is in the overtaking lane, by enforcing a minimum lateral distance between the two vehicles $s^{y}(t) \geq s_{Y_{c}}$, where $s_{Y_{c}}=2d_{Y_{0}}$. 
The horizontal green segment further intercepts with the blue and purple segments at points ($s_{X_{b}}, s_{Y_{b}}$) and ($s_{X_{d}}, s_{Y_{d}}$), respectively, such that the Euclidean distance between the two vehicle geometric centers when the EV is at either ($s_{X_{b}}, s_{Y_{b}}$) or ($s_{X_{d}}, s_{Y_{d}}$) equals the minimum safe distance, $2r$, according to the arc-polygons of the two cars merely touching.
Therefore, the longitudinal coordinates $s_{X_{b}}=-s_{X_{d}}=-\sqrt{(2r)^2-s_{Y_{c}}^2}$ and the lateral coordinates $s_{Y_b}=s_{Y_{c}}=s_{Y_{d}}$.

After determining the coordinates of all the relevant line segment points, the constraints to avoid collisions with the OV can be uniformly represented as
\begin{equation}\label{eq:col_avod}
    \begin{aligned}
    k_{p}(t)s^{x}(t)+b_{p}(t)&\leq s^{y}(t), \qquad \forall \; p \in \{1,2,3 \}, 
    \end{aligned}
\end{equation}
where $p$ is the constraint index inferred from the blue ($p=1$), green ($p=2$), and purple ($p=3$) boundaries, respectively, with 
\begin{subequations} \label{eq:col_avod_numerical_expressions}
    \begin{align}
    k_{1}(t)&=\frac{s_{Y_b}-s_{Y_a}}{s_{X_b}-s_{X_a}(t)}, & b_{1}(t)&=s_{Y_a}-\frac{s_{X_a}(t)(s_{Y_b}-s_{Y_a})}{s_{X_b}-s_{X_a}(t)}, 
    \label{eq:col_avod_numerical_expressions1}\\
    k_{2}(t)&=0, & b_{2}(t)&=s_{Y_c}, 
    \label{eq:col_avod_numerical_expressions2}\\
    k_{3}(t)&=\frac{s_{Y_e}-s_{Y_d}}{s_{X_e}(t)-s_{X_d}}, & b_{3}(t)&= s_{Y_d}-\frac{s_{X_d}(s_{Y_e}-s_{Y_d})}{s_{X_e}(t)-s_{X_d}}. 
    \label{eq:col_avod_numerical_expressions3}
    \end{align}
\end{subequations} 

\begin{figure*}[h!]
\centering
\includegraphics[width=\textwidth]{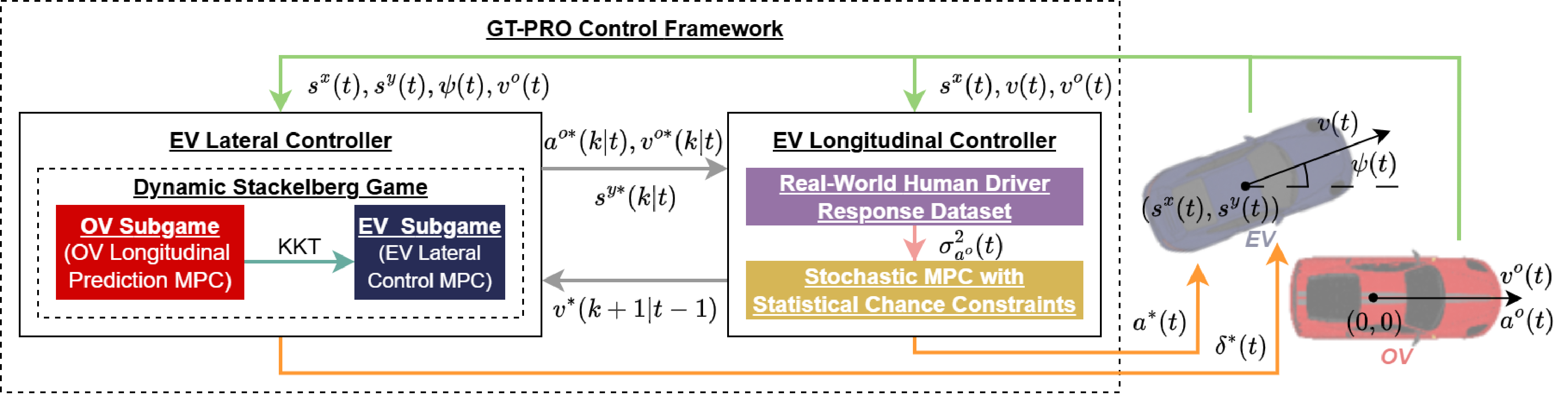}
\caption{Overall architecture of the proposed GT-PRO strategy. The EV lateral and longitudinal controllers determine the optimal EV front wheel steering angle ($\delta^{*}(t)$) and longitudinal acceleration ($a^{*}(t)$), respectively, as control inputs denoted by orange arrows. The green arrows represent the measurements of the current EV and OV states. 
$v(t)$ and $\psi(t)$ are the longitudinal velocity and heading of the EV, $v^{o}(t)$ and $a^{o}(t)$ are the longitudinal velocity and acceleration of the OV, and $s^{x}(t)$ and $s^{y}(t)$ are the longitudinal and lateral positions of the EV, respectively. All positions are given with respect to the $X$-$Y$ coordinate system (as defined in Fig. \ref{fig:bicycle_model_arcpolyton}) with the origin at the OV geometric center $(0,0)$.
The EV lateral controller solves a dynamic Stackelberg game-based MPC problem, which is formulated as a constrained bilevel optimization. The lower-level OV subgame is converted into the KKT conditions of the upper-level EV subgame.
The EV longitudinal controller analyzes a real-world human driver response dataset to interpret the variance of human longitudinal driving behavior ($\sigma_{a^{o}}^{2}(t)$), which is used to formulate statistical chance constraints of the stochastic MPC in the EV longitudinal controller. 
Gray arrows represent internally shared sequences: 
$a^{o*}(k|t)$, $v^{o*}(k|t)$, and $s^{y*}(k|t)$ are the predicted OV longitudinal acceleration and velocity, and the optimal EV lateral position sequences shared by the EV lateral controller, respectively; $v^{*}(k+1|t-1)$ is the optimal EV longitudinal velocity sequences shared by the EV longitudinal controller, generated in the previous iteration.}
\label{fig:overall_block_diagram}
\end{figure*}
Furthermore, the yellow segments correspond to the collision avoidance constraints implemented to prevent collisions with roadside obstacles, ensuring that the EV maintains a minimum distance of at least $d_{Y_{0}}$ from the road boundaries, such that
\begin{equation} \label{eq:col_avod_roadside}
    -\frac{W_l}{2}+d_{Y0} \leq s^{y}(t) \leq \frac{3W_{l}}{2}-d_{Y0}.
\end{equation}

\section{Controller Design}
\label{sec:control design}
The proposed GT-PRO framework is composed of two modules responsible for lateral and longitudinal control of the EV, respectively, as shown in Fig. \ref{fig:overall_block_diagram}. 
Since the lateral and longitudinal dynamics of the EV are coupled and nonlinear, two approximations are adopted to linearize and decouple these dynamics, which can then be used in the design of the two control modules. First, a small-angle approximation of the EV steering and heading angles is adopted, such that the EV dynamics in \eqref{eq:vehicle_bicycle_model1}-\eqref{eq:vehicle_bicycle_model3} become
\begin{subequations}\label{eq:vehicle_bicycle_models_linearised}
    \begin{align}
         s^{x}(t+1)&=s^{x}(t)+\left( v(t)-v^{o}(t)\right)\Delta T, \label{eq:vehicle_bicycle_model1_linearised}\\
         s^{y}(t+1)&=s^{y}(t)+v(t)\psi(t)  \Delta T,\label{eq:vehicle_bicycle_model2_linearised}\\
         \psi(t+1)&=\psi(t)+\frac{v(t)\delta(t)}{l} \Delta T. \label{eq:vehicle_bicycle_model3_linearised}
    \end{align}
\end{subequations}
Secondly, the lateral control module considers the longitudinal motion (the EV longitudinal velocity, $v(t)$) to be given exogenously by the longitudinal control module, and vice versa the longitudinal control module considers the lateral motion (the EV lateral position, $s^{y}(t)$) to be given exogenously by the lateral control module. Thus, for example, within the lateral control module the nonlinear terms $v(t)\psi(t)$ in \eqref{eq:vehicle_bicycle_model2_linearised} and $v(t)\delta(t)$ in \eqref{eq:vehicle_bicycle_model3_linearised} become products of a known function of time (exogenous input $v(t)$ from the longitudinal controller) and a state or control input, therefore making the approximated system dynamics linear parameter varying.
Furthermore, although the EV lateral and longitudinal dynamics are decoupled in the two modules in this way, the two control modules are coupled and work collaboratively, as in a distributed control framework \cite{Zheng2017,Dunbar2006}, by sharing with each other OV and EV state prediction sequences from their respective optimization solutions. This distributed framework design avoids the inaccuracy caused by approximating time-varying variables as constant parameters, as done, for example, in \cite{Lizhuoren2024}.

With reference to Fig. \ref{fig:overall_block_diagram}, the more precise details of the developed overall scheme are as follows.
At each sampling time $t$, where $k\in \{0, 1, \dots, N-1\}$ denotes the step index within a prediction horizon of length $N$: a) the EV lateral controller receives from the EV and OV state measurements and from the longitudinal controller the optimal EV longitudinal velocity trajectory, $v^{*}(k+1|t-1)$, from the previous $(t-1)$-th step, to facilitate computations at the $t$-th step, and b) the EV lateral control module solves a dynamic Stackelberg game-based MPC, determining the optimal front wheel steering angle sequence of the EV ($\delta^{*}(k|t)$) and simultaneously predicting the longitudinal acceleration behavior of the OV ($a^{o*}(k|t)$).

After the lateral control module performs its calculations at the $t$-th step, it passes to the longitudinal control module, which involves a stochastic MPC scheme, the predicted OV longitudinal acceleration ($a^{o*}(k|t)$), velocity ($v^{o*}(k|t)$), and the optimal EV lateral trajectory ($s^{y*}(k|t)$). 
The longitudinal controller deals with human driver uncertainties that are not considered in the lateral control module by analyzing a real-world human driver response dataset \cite{highDdataset}. Statistical analysis of this dataset numerically characterizes the Gaussian distribution variance ($\sigma^{2}_{a^o}(t)$) of human driver behavior uncertainty, which is utilized to implement a chance constraint of the stochastic MPC. Thus, the optimal longitudinal driving effort of the EV ($a^{*}(k|t)$) is determined by the stochastic MPC at the $t$-th step. When all the calculations from both control modules are completed, the first element of each control sequence, $\delta^{*}(t)$ and $a^{*}(t)$, is applied to the EV and the whole procedure described for the two control modules is repeated at the next sampling time in a receding horizon manner for $N_{T}$ iterations.

For clarity, the following notation is introduced, which is used in the MPC formulations of this work described subsequently, where $\xi$ is a generic example variable:
\begin{itemize}
    \item $\xi(t)$ denotes the actual system state/input/auxiliary variable at the sampling time $t$.
    \item $\xi(k|t)$ denotes the predicted state/input/auxiliary variable at step $k$ given information at time $t$. $\check{\xi}(k|t)$ and $\hat{\xi}(k|t)$ denote the state/input/auxiliary variable $\xi(k|t)$ that is used in the internal model of the lateral and longitudinal control modules, respectively.
    \item $\xi^{*}(k|t)$ denotes the optimal state/input/auxiliary variable at step $k$ given information at time $t$, determined by the associated control module.
    \item $\boldsymbol{\xi}$ denotes the stacked state/input/auxiliary vector of $\xi(k|t)$ over the prediction horizon $[t, t+N]$.
\end{itemize}

Section \ref{sec:lateral_controller} and Section \ref{sec:longitudinal_controller} describe, respectively, the longitudinal and lateral control modules shown in Fig. \ref{fig:overall_block_diagram}. 

\subsection{Stackelberg game-based lateral control module}
\label{sec:lateral_controller}
\subsubsection{Formulation of the Stackelberg game-based interaction}
\label{sec:stackelberg_game}
A Stackelberg game-based approach is employed to model the interactive behavior of the OV when the 
EV aims to overtake it. 
Within the Stackelberg game structure, the EV acts first by deciding its lateral control effort $u^{y}(t)$ based on its objective function $J^{y}(t)$, followed by the OV, which is assumed to adjust its action $u^{o}(t)$ to minimize its own cost $J^{o}(t)$ in response to the EV’s move. Therefore,
\begin{equation}
    \begin{aligned}
        {u}^{y*}(t)&\!=\!\argmin\limits_{{u}^{y}\!(t)\in\mathcal{U}^{y}(u^{o}\!(t))}\!\left(\min\limits_{{u}^{o}\!(t)\in\mathcal{U}^{o*}(u^{y}(t))} J^{y}\!\left(x^{y}\!(t),{u}^{y}\!(t)\right)\!  \right), \label{eq:stackelberg_formulation1}\\
    \end{aligned}
\end{equation} 
where $x^{y}(t)$ is the EV state vector, and $\mathcal{U}^{y}(u^{o}(t))$ is the EV feasible space determined by the OV action, while the optimal feasible space of the OV $\mathcal{U}^{o*}\left({u}^{y}(t)\right)$ is determined as
\begin{equation}
    \begin{aligned}
        \mathcal{U}^{o*}\left({u}^{y}\!(t)\right)&=\Bigl\{{u}^{o*}(t)\in \mathcal{U}^{o}(u^{y}(t)): J^{o}\left(x^{o}(t),{u}^{o*}(t)\right) \\
         &\leq J^{o}\left(x^{o}(t),{u}^{o}(t)\right), \forall {u}^{y}(t)\in \mathcal{U}^{y}\left(u^{o}(t)\right)\Bigr\}, \label{eq:stackelberg_formulation2}
    \end{aligned}
\end{equation}
where $x^{o}(t)$ is the OV state vector, and $\mathcal{U}^{o}(u^{y}(t))$ is the feasible space of the OV determined by the action of the EV. 
The game is resolved using backward induction to obtain the optimal solution for each subgame, which naturally corresponds to a bilevel optimization structure: the OV subgame (lower-level optimization) is solved first to determine its best response function for all given EV action; this best response is embedded into the EV subgame (upper-level optimization), which is then solved to minimize the EV cost. 

\subsubsection{Formulation of the OV subgame}
\label{sec:FHOCP_OV}
The OV subgame is formulated as a finite-horizon optimal control problem (FHOCP).
By collecting dynamic equations in \eqref{eq:vehicle_bicycle_model1_linearised} and \eqref{eq:vehicle_bicycle_model5}, the state equation of the OV subgame is 
\begin{equation}
    \begin{aligned}
        \!\!\!x^{o}(k+1|t)&=A^{o}x^{o}(k|t)+B^{o}u^{o}(k|t)+E^{o}u^{o}_e(k|t),\label{eq:ov_state_space_x} \\
        \!\!\!A^{o}&=\begin{bmatrix}
        1&-\Delta T\\0&1\end{bmatrix}, \quad
        B^{o}=\begin{bmatrix}
         0\\\Delta T\end{bmatrix},\quad 
         E^{o}=\begin{bmatrix}\Delta T\\ 0\end{bmatrix},
    \end{aligned}
\end{equation}
where the state vector $x^{o}\!(\!k|t\!)\!=\![\check{s}^{x}\!(\!k|t\!),\check{v}^{o}\!(\!k|t\!)]^{\top}$, the control input of the OV subgame $u^{o}\!(\!k|t\!)\!=\![\check{a}^{o}\!(\!k|t\!)]$, and the exogenous input $u^{o}_{e}(k|t)=[v^{*}(k\!+\!1|t\!-\!1)]$ is the shifted optimal EV longitudinal velocity sequence generated by the longitudinal controller at the previous iteration.

In addition to the collision avoidance constraint \eqref{eq:col_avod}, the state and control variables of the OV subgame are subject to boundary constraints. To express these compactly, we introduce the constraint vector $f^{o}(k|t)=[\check{v}^{o}(k|t),\check{a}^{o}(k|t)]^{\top}$, which satisfies 
\begin{equation}\label{eq:lat_constraint_OV}
    \begin{aligned}
        \underline{f}^{o}\leq f^{o}(k|t)\leq \overline{f}^{o},
    \end{aligned}
\end{equation}
where $\underline{f}^{o}=[\underline{v}^{o},\underline{a}]^{\top}$ and  $\overline{f}^{o}=[\overline{v}^{o},\overline{a}]^{\top}$ denote the lower and upper bounds of the OV speed and acceleration, respectively.

The objective of the OV subgame is to address three aspects: speed, safety, and comfort. The OV is assumed to have a general preference for maintaining its current speed ($v^{o}(t)$) when being overtaken. In addition, due to the safety concern, the OV can choose to expand the headway space by tracking for $\check{s}^x(k|t)$ a target longitudinal distance between vehicles $\tilde{s}^{x}(t)=d_{X_{0}}+v^{o}(t)\tilde{t}$, which is constant within a horizon, with $\tilde{t}$ the vehicle target headway time. This headway expanding objective is active once the EV is marginally ahead of the OV and until the headway distance already exceeds $\tilde{s}^{x}(t)$, i.e.,
\begin{equation}\label{eq:lat_headwayexpanding}
    \begin{aligned}
        s_{X_{d}}\leq s^{x}(t)\leq \tilde{s}^{x}(t). 
    \end{aligned}
\end{equation}
When the OV adjusts its speed, it also aims to avoid uncomfortable hard braking or acceleration by penalizing its acceleration ($\check{a}^{o}(k|t)$). Therefore, cost and cost reference vectors are defined as $z^{o}(k|t)=[\check{s}^{x}(k|t),\check{v}^{o}(k|t),\check{a}^{o}(k|t)]^{\top}$ and $\tilde{z}^{o}(k|t)=[\tilde{s}^{x}(t),v^{o}(t),0]^{\top}$, respectively, and the FHOCP of the OV subgame is formulated as 
\begin{subequations}\label{eq:ov_OCP1}
\begin{align}
\!\!\!\!\!\!\min\limits_{u^{o}(k|t)}\quad& J^{o}\!(t)\!\!=\!\!\!\sum_{k=0}^{N}\!\!\left(z^{o}\!(k|t)\!-\!\tilde{z}^{o}\!(k|t)\!\right)^{\!\!\top\!}\!Q^{o}(t)\!\left(z^{o}\!(k|t)\!-\!\tilde{z}^{o}\!(k|t)\!\right),\\
\!\!\!\!\!\!\textbf{s.t. } & \check{k}_{p}(k|t)\check{s}^{x}(k|t)+\check{b}_{p}(k|t)\leq \check{s}^{y}(k|t), \label{eq:ov_OCP1_coll} \\ 
&
\eqref{eq:col_avod_numerical_expressions},\eqref{eq:ov_state_space_x},\eqref{eq:lat_constraint_OV}, \eqref{eq:lat_headwayexpanding}\; \forall \;  p\! \in\! \{1,2,3 \}, \;  k\!\in\! \mathbb{N}_{[0, N-1]},\\
\!\!\!\!\!\!\textbf{given: }& x^{o}(t)\!=\![s^{x}(t),v^{o}(t)]^{\top}, \\
\!\!\!\!\!\!& v^{*}(k+1|t-1), v^{o*}(k+1|t-1), \forall \; k\!\in\! \mathbb{N}_{[0, N-1]},
\end{align}
\end{subequations}
where $Q^{o}(t)\!\in\! \mathbb{S}^{3}_{\succeq}$ is the weighting matrix that tunes the multi-objective cost function. The weighting parameters are chosen by trial-and-error such that the predicted OV response is aligned with the most common human driver responses from the real-world dataset \cite{highDdataset}. The time-varying $Q^{o}(t)$ determines if the headway expanding objective is in effect, such that when \eqref{eq:lat_headwayexpanding} is violated, the corresponding element of this matrix is set to zero.
$\check{s}^{y}(k|t)$ in \eqref{eq:ov_OCP1_coll} is the EV lateral position managed by the EV subgame that will be introduced in Section \ref{sec:FHOCP_EV}.
When $p=2$, the sequences of $\check{k}_{p}(k|t)$ and $\check{b}_{p}(k|t)$ are constant according to \eqref{eq:col_avod_numerical_expressions2}. On the contrary, for $p=1$, $\check{k}_{p}(k|t)$ and $\check{b}_{p}(k|t)$ are constructed by following \eqref{eq:col_avod_numerical_expressions1} where $\check{s}_{X_a}(k|t)$ is calculated after substituting $v^{*}(k+1|t-1)$ from the longitudinal controller into \eqref{eq:coordinates_1}.
Analogously, for $p=3$, $\check{k}_{p}(k|t)$ and $\check{b}_{p}(k|t)$ are determined by \eqref{eq:col_avod_numerical_expressions3} where $\check{s}_{X_e}(k|t)$ is obtained after substituting $v^{o*}(k+1|t-1)$ from the last iteration of the lateral controller into \eqref{eq:coordinates_2}. 

\subsubsection{Formulation of the EV subgame}
\label{sec:FHOCP_EV}
The EV determines its optimal steering angle by solving the EV subgame. By collecting \eqref{eq:vehicle_bicycle_model2_linearised} and \eqref{eq:vehicle_bicycle_model3_linearised}, the state vector of the EV subgame is $x^{y}(k|t)=[\check{s}^{y}(k|t),\check{\psi}(k|t)]^{\top}$ and the control vector is $u^{y}(k|t)=[\check{\delta}(k|t)]$. The linearized state-space equation of the lateral dynamics is summarized as
\begin{equation} 
    \begin{aligned} 
    &\!\!\!\!\!\!x^{y}(k+1|t)=A^{y}(k|t)x^{y}(k|t)+B^{y}(k|t)u^{y}(k|t),\label{eq:lat_state_space_x} \\ &\!\!\!\!\!\!A^{y}\!(k|t)\!=\!\!\begin{bmatrix} 1&\!\!v^{*}(k\!+\!1|t\!-\!1)\Delta T\\0&\!\!1\end{bmatrix}, B^{y}\!(k|t)\!=\!\!\begin{bmatrix} 0\\\frac{v^{\!*}(k+1|t-1)\Delta T}{l}\end{bmatrix}\!,
    \end{aligned} 
\end{equation}
where $A^{y}(k|t)$ and $B^{y}(k|t)$ are time-variant matrices that depend linearly on $v^{*}(k+1|t-1)$. Moreover, the EV subgame is subject to associated constraints, including the collision avoidance constraints \eqref{eq:col_avod}-\eqref{eq:col_avod_roadside}, the heading angle limit ($\underline{\psi}$ and $\overline{\psi}$), and the steering angle limit ($\underline{\delta}$ and $\overline{\delta}$)
\begin{subequations} \label{eq:lat_constraint_EV_angles}
    \begin{align}
        \underline{\psi}&\leq {\check{\psi}}(k|t)\leq \overline{\psi}, \label{eq:lat_constraint_EV_psi} \\
        \underline{\delta}&\leq {\check{\delta}}(k|t)\leq \overline{\delta}. \label{eq:lat_constraint_EV_delta}
    \end{align}
\end{subequations}
The implementations of \eqref{eq:lat_constraint_EV_angles} also ensure that the small-angle approximation remains valid, so that the approximations in \eqref{eq:vehicle_bicycle_models_linearised} are accurate. For the convenience of further implementation, \eqref{eq:col_avod_roadside} and \eqref{eq:lat_constraint_EV_angles} can be compactly summarized as 
\begin{equation}\label{eq:lat_constraint_EV}
    \begin{aligned}
        \underline{f}^{y}\leq f^{y}(k|t)\leq \overline{f}^{y},
    \end{aligned}
\end{equation}
where $f^{y}(k|t)=[\check{s}^{y}(k|t),\check{\psi}(k|t),\check{\delta}(k|t)]^{\top}$ is the constraint vector, $\underline{f}^{y}=[\underline{s}^{y},\underline{\psi},\underline{\delta}]^{\top}$, and $\overline{f}^{y}=[\overline{s}^{y},\overline{\psi},\overline{\delta}]^{\top}$. 

The optimization objective of the EV subgame includes three terms. In order to execute the pulling-out and cutting-in maneuvers, the target lateral position of the EV ($\tilde{s}^{y}(t)$) is set to the middle of the overtaking lane while the EV passes next to the OV, and to the middle of the initial lane otherwise, to avoid over-occupancy of the overtaking lane:
\begin{equation} \label{eq:lat_objective_function_reference}
    \begin{aligned}
        \tilde{s}^{y}(t)\!=\! \left\{\!\!\!\!
 \begin{array}{ll}
   \displaystyle
    W_l, &  s_{X_{b}}\leq s^{x}(t)\leq s_{X_{d}}, \\
    \\
    \displaystyle  0, &   \left( s^{x}(t)< s_{X_{b}} \right) \lor \left(s^{x}(t)>s_{X_{d}} \right).\\
 \end{array}\right.    
    \end{aligned}
\end{equation}
During lane changes before and after overtaking, it is also desired to keep the vehicle heading angle $\check{\psi}(k|t)$ low to improve safety. In addition, according to the common approach in the literature \cite{deWinkel2023}, comfort during overtaking can be quantified by solely investigating the EV lateral acceleration, $a^{y}(t)$, which can be approximated following a small side slip angle assumption by ${a}^{y}(t)\!\approx\! \frac{v(t)^2}{l}{\delta}(t)$ \cite{Matute2019}. 
Furthermore, minimizing lateral acceleration is essential to prevent exceeding the acceleration ellipse limits, which would compromise the accuracy of the decoupled control scheme proposed here, and to maintain vehicle stability, thereby reducing the risk of drifting during the overtaking maneuver.    
Since the EV longitudinal velocity is not controlled within the lateral controller, it is reasonable to penalize the $L_{2}$-norm of $\check{\delta}(k|t)$ to minimize the lateral acceleration magnitude, according to the acceleration approximation above. 
In order to capture the above objectives, a cost vector $z^{y}(k|t)=[\check{s}^{y}(k|t),\check{\psi}(k|t),\check{\delta}(k|t)]^{\top}$ and an associated reference vector $\tilde{z}^{y}(k|t)=[\tilde{s}^{y}(t),0,0]^{\top}$ are defined. Therefore, the EV subgame can be formulated as an FHOCP that determines the optimal steering angle at time $t$ as
\begin{subequations}\label{eq:ev_OCP1}
\begin{align}
\!\!\!\min\limits_{u^{y}(k|t)}\quad & J^{y}\!(t)\!\!=\!\!\!\sum_{k=0}^{N}\!\!\left(z^{y}\!(k|t)\!-\!\tilde{z}^{y}\!(k|t)\!\right)^{\!\!\top\!}\!Q^{y}(t)\!\left(z^{y}\!(k|t)\!-\!\tilde{z}^{y}\!(k|t)\!\right), \label{eq:ev_OCP1_cost}\\
\!\!\!\textbf{s.t. } & \check{k}_{p}(k|t)\check{s}^{x}(k|t)+\check{b}_{p}(k|t)\leq \check{s}^{y}(k|t), \label{eq:ev_OCP1_coll}\\ 
&
\eqref{eq:col_avod_numerical_expressions},\eqref{eq:lat_state_space_x},\eqref{eq:lat_constraint_EV},\eqref{eq:lat_objective_function_reference}, \forall \;  p\!\in\! \{1,2,3\}, k\!\in\! \mathbb{N}_{[0, N-1]},\\
\!\!\!\textbf{given: }& x^{y}(t)\!=\![s^{y}(t),\psi(t)]^{\top},\\
\!\!\!&v^{*}(k+1|t-1), v^{o*}(k+1|t-1), \forall \; k\!\in\! \mathbb{N}_{[0, N-1]},
\end{align}
\end{subequations}
where $Q^{y}(t) \in \mathbb{S}^{3}_{\succ}$ is the weighting matrix. 
Similar to \eqref{eq:ov_OCP1}, $\check{k}_{p}(k|t)$, $\check{b}_{p}(k|t)$ together with $\check{s}^{x}(k|t)$ predicted by the OV subgame in \eqref{eq:ov_OCP1} contribute the lower boundary of $\check{s}^{y}(k|t)$ in \eqref{eq:ev_OCP1_coll}.

The dynamic Stackelberg game consisting of the lower-level (follower) OV subgame problem \eqref{eq:ov_OCP1} and the upper-level (leader) EV subgame problem \eqref{eq:ev_OCP1} is a bilevel optimization. To find the numerical solution of the bilevel optimization, both the leader and follower problems are expressed in condensed (stacked) MPC form over the prediction horizon, where the following intermediate equations are used:
\begin{subequations} \label{eq:KKT_intermediate_expressions}
    \begin{align}
        & \boldsymbol{x}^{y}=\tilde{A}^{y}x^{y}(t)+\tilde{B}^{y}\boldsymbol{u}^{y}, \label{eq:KKT_intermediate_expression_x}\\
        & \boldsymbol{f}^{y}=\tilde{C}_{f}^{y}x^{y}(t)+\tilde{D}_{f}^{y}\boldsymbol{u}^{y}, \label{eq:KKT_intermediate_expression_f}\\
        & \boldsymbol{z}^{y}=\tilde{C}_{z}^{y}x^{y}(t)+\tilde{D}_{z}^{y}\boldsymbol{u}^{y},\label{eq:KKT_intermediate_expression_z}\\
        & \boldsymbol{x}^{o}=\tilde{A}^{o}x^{o}(t)+\tilde{B}^{o}\boldsymbol{u}^{o}+\tilde{E} ^{o}\boldsymbol{u}^{o}_{e},\label{eq:KKT_intermediate_expression_xo}\\
        & \boldsymbol{f}^{o}=\tilde{C}_{f}^{o}x^{o}(t)+\tilde{D}_{f}^{o}\boldsymbol{u}^{o}+\tilde{E}_{f}^{o}\boldsymbol{u}^{o}_{e}, \label{eq:KKT_intermediate_expression_fo}\\
        & \boldsymbol{z}^{o}=\tilde{C}_{z}^{o}x^{o}(t)+\tilde{D}_{z}^{o}\boldsymbol{u}^{o}+\tilde{E}_{z}^{o}\boldsymbol{u}^{o}_{e}, \label{eq:KKT_intermediate_expression_zo}
    \end{align}
\end{subequations}
where \eqref{eq:KKT_intermediate_expression_x} and \eqref{eq:KKT_intermediate_expression_xo} define the stacked state vectors $\boldsymbol{x}^{y}\!\in\! \mathbb{R}^{2(N+1)}$ and $\boldsymbol{x}^{o}\!\in\! \mathbb{R}^{2(N+1)}$ by iterating the state-space dynamics in \eqref{eq:lat_state_space_x} and \eqref{eq:ov_state_space_x}, respectively. $\boldsymbol{u}^{y} \!\in\! \mathbb{R}^{N}$ and $\boldsymbol{u}^{o} \!\in\! \mathbb{R}^{N}$ are stacked input vectors of $u^{y}(k|t)$ and $u^{o}(k|t)$. The matrices $\tilde{A}^{y}$, $\tilde{B}^{y}$, $\tilde{A}^{o}$, $\tilde{B}^{o}$, and $\tilde{E}^{o}$ are the associated condensed system matrices.
Similarly, $\boldsymbol{f}^{y}\!\in\! \mathbb{R}^{3(N+1)}$ and $\boldsymbol{z}^{y}\!\in\! \mathbb{R}^{3(N+1)}$ are the stacked vectors of $f^{y}(k|t)$ and $z^{y}(k|t)$, with $\tilde{C}_{f}^{y}$, $\tilde{D}_{f}^{y}$, $\tilde{C}_{z}^{y}$, and $\tilde{D}_{z}^{y}$ denoting the corresponding condensed coefficient matrices collecting linear combination of state and input variables. Analogously, $\boldsymbol{f}^{o}\!\in\! \mathbb{R}^{2(N+1)}$ and $\boldsymbol{z}^{o}\!\in\! \mathbb{R}^{3(N+1)}$ are stacked vectors of $f^{o}(k|t)$ and $z^{o}(k|t)$, associated with condensed matrices $\tilde{C}_{f}^{o}$, $\tilde{D}_{f}^{o}$, $\tilde{E}_{f}^{o}$, $\tilde{C}_{z}^{o}$, $\tilde{D}_{z}^{o}$, and $\tilde{E}_{z}^{o}$.  
Furthermore, the lower-level problem is replaced by its KKT conditions, comprising stationarity, primal feasibility, dual feasibility, and complementary slackness, which are incorporated as constraints into the upper-level problem \cite{Zhang2024}. This yields a single-level Mathematical Program with Equilibrium Constraints (MPEC) \cite{Luo1996} formulation presented below, in which the EV (leader's) objective is optimized subject to both the upper-level EV’s own constraints and the lower-level OV’s equilibrium KKT constraints, while preserving the original Stackelberg structure:
\begin{subequations} \label{eq:KKT}
    \begin{align}
        \min\limits_{\substack{\boldsymbol{u}^{y}, \boldsymbol{u}^{o}, \underline{\boldsymbol{\lambda}}, \overline{\boldsymbol{\lambda}}}}\quad & {J}^{y}(t)=\left(\boldsymbol{z}^{y}-\tilde{\boldsymbol{z}}^{y}\right)^{\top}\boldsymbol{Q}^{y}\left(\boldsymbol{z}^{y}-\tilde{\boldsymbol{z}}^{y}\right), \label{eq:KKT_cost} \\
\!\!\!\textbf{s.t. } & \text{Intermediate expressions:} \quad \eqref{eq:KKT_intermediate_expressions}\nonumber\\
& \text{Upper-level inequality constraints:} \nonumber\\
& \quad \boldsymbol{f}^{y} - \overline{\boldsymbol{f}}^{y} \leq 0,\quad  \underline{\boldsymbol{f}}^{y}- \boldsymbol{f}^{y} \leq 0, \label{eq:KKT_upper_constraint1}\\
& \quad \left( \check{\boldsymbol{k}}_{p}\mathbf{I}_{1}\boldsymbol{x}^{o}+\check{\boldsymbol{b}}_{p}\right) - \mathbf{I}_{1}\boldsymbol{x}^{y} \leq 0, \; \forall \;  p\!\in\! \{1,2,3 \}, \label{eq:KKT_upper_constraint2}\\
& \text{Lower-level KKT conditions:} \nonumber\\
& \quad \text{Stationarity:} \nonumber\\
& \quad 2\tilde{D}_{z}^{o\top}\boldsymbol{Q}^{o}\left( \boldsymbol{z}^{o}- \tilde{\boldsymbol{z}}^{o} \right)+ \tilde{D}_{f}^{o\top}\left( \overline{\boldsymbol{\lambda}}-\underline{\boldsymbol{\lambda}} \right) = 0, \label{eq:KKT_stationary}\\
& \quad \text{Primal feasibility:} \nonumber\\
& \quad \boldsymbol{f}^{o} - \overline{\boldsymbol{f}}^{o} \leq 0 , \quad \underline{\boldsymbol{f}}^{o}-\boldsymbol{f}^{o} \leq 0, \label{eq:KKT_primal_feasibility}\\
& \quad\text{Dual feasibility:} \nonumber\\
& \quad\overline{\boldsymbol{\lambda}} \geq 0, \quad \underline{\boldsymbol{\lambda}} \geq 0, \label{eq:KKT_dual_feasibility} \\
& \quad \text{Complementary slackness:} \nonumber\\
& \quad \overline{\boldsymbol{\lambda}} \circ \left( \boldsymbol{f}^{o} - \overline{\boldsymbol{f}}^{o} \right) =0,\quad \underline{\boldsymbol{\lambda}} \circ \left( \underline{\boldsymbol{f}}^{o}-\boldsymbol{f}^{o} \right) =0, \label{eq:KKT_slackness} \\
\!\!\!\textbf{given: }& x^{o}(t)\!=\![s^{x}(t), v^{o}(t)]^{\top},  \quad x^{y}(t)\!=\![s^{y}(t),\psi(t)]^{\top},\nonumber \\
& v^{*}(k+1|t-1), v^{o*}(k+1|t-1), \forall \; k\!\in\! \mathbb{N}_{[0, N-1]}, \nonumber
    \end{align}
\end{subequations}
where $\mathbf{I}_{1}=I^{N+1}\otimes [1,0]$, $I^{N+1}$ is an identity matrix of size $N+1$, $\otimes$ is the Kronecker tensor product, and $\circ$ is the Hadamard (element-wise) product; 
$\underline{\boldsymbol{\lambda}}, \overline{\boldsymbol{\lambda}} \!\in\! \mathbb{R}^{2(N+1)}$ are Lagrange multipliers; 
$\boldsymbol{Q}^{y}\!\in\! \mathbb{S}^{3(N+1)}_{\succ}$ and $\boldsymbol{Q}^{o}\!\in\! \mathbb{S}^{3(N+1)}_{\succeq}$ are stacked weighting matrices of $Q^{y}(t)$ and $Q^{o}(t)$, respectively;
$\tilde{\boldsymbol{z}}^{y}\!\in\! \mathbb{R}^{3(N+1)}$, $\underline{\boldsymbol{f}}^{y} \!\in\! \mathbb{R}^{3(N+1)}$, $\overline{\boldsymbol{f}}^{y} \!\in\! \mathbb{R}^{3(N+1)}$, $\tilde{\boldsymbol{z}}^{o} \!\in\! \mathbb{R}^{3(N+1)}$, $\underline{\boldsymbol{f}}^{o} \!\in\! \mathbb{R}^{2(N+1)}$, and $\overline{\boldsymbol{f}}^{o} \!\in\! \mathbb{R}^{2(N+1)}$ are stacked vectors of $\tilde{z}^{y}(k|t)$, $\underline{f}^{y}$, $\overline{f}^{y}$, $\underline{f}^{o}$, and $\overline{f}^{o}$, respectively;
$\check{\boldsymbol{k}}_{p}\!\in\! \mathbb{D}^{N+1}$ and $\check{\boldsymbol{b}}_{p}\!\in\! \mathbb{R}^{N+1}$ are stacked diagonal matrix of $\check{k}_{p}(k|t)$ and vector of $\check{b}_{p}(k|t)$, respectively;
\eqref{eq:KKT_cost}-\eqref{eq:KKT_upper_constraint2} are condensed forms of \eqref{eq:ev_OCP1_cost}, \eqref{eq:lat_constraint_EV}, and \eqref{eq:ev_OCP1_coll}, respectively, and \eqref{eq:KKT_stationary}-\eqref{eq:KKT_slackness} are equivalent KKT conditions of the OV subgame \eqref{eq:ov_OCP1}.

\subsection{Statistical MPC-based longitudinal control module}
\label{sec:longitudinal_controller}
\subsubsection{Statistical Human Driver Uncertainty Modeling}
\label{sec:dataset}
The model used by the stochastic MPC of the longitudinal control module includes the dynamic equations \eqref{eq:vehicle_bicycle_model1_linearised},\eqref{eq:vehicle_bicycle_model4}, and \eqref{eq:vehicle_bicycle_model5}, for which the state-space vector $x^{x}(k|t)=[\hat{s}^{x}(k|t),\hat{v}(k|t),\hat{v}^{o}(k|t)]^{\top}$ and the control variable $u^{x}(k|t)=[\hat{a}(k|t)]$ are defined, such that the state-space equation is written as
\begin{equation}
    \begin{aligned}
        &x^{x}(k+1|t)=A^{x}x^{x}(k|t)+B^{x}u^{x}(k|t)+E^{x}w^{x}(k|t),\label{eq:lon_state_space_x} \\
        &A^{x}\!=\!\begin{bmatrix}1&\Delta T&-\Delta T\\0&1&0\\0&0&1\end{bmatrix},  B^{x}\!=\!\begin{bmatrix}0\\\Delta T\\0\end{bmatrix},
        E^{x}\!=\!\begin{bmatrix}0\\0\\\Delta T\end{bmatrix},
    \end{aligned}
\end{equation}
where $A^{x}$, $B^{x}$, and $E^{x}$ are time-invariant matrices, and $w^{x}(k|t)$ represents the uncertain OV acceleration as a disturbance. 
Although the lateral controller introduced in Section \ref{sec:lateral_controller} predicts the OV acceleration through the game-theoretic strategy, a mismatch between the actual and predicted OV accelerations still exists due to the human driver behavior uncertainty. To account for this deviation, which can be modeled as a stochastic uncertainty, the additive disturbance $w^{x}(k|t)\sim \mathcal{N}({a}^{o*}(k|t),\sigma^{2}_{a^{o}}(k|t))$ is introduced in the longitudinal control module stochastic MPC as a normally distributed random variable with a mean of the lateral control module game-predicted OV acceleration $({a}^{o*}(k|t))$ and a variance $\sigma^{2}_{a^{o}}(k|t)$. This variance is fixed over the prediction horizon ($\sigma^{2}_{a^{o}}(k|t)=\sigma^{2}_{a^{o}}(t)$) and is calculated from a real-world human driver response dataset, as explained below. 

\begin{figure}[t!]
\centering
    \begin{subfigure}{\columnwidth}
         \centering
         \includegraphics[width=\columnwidth]{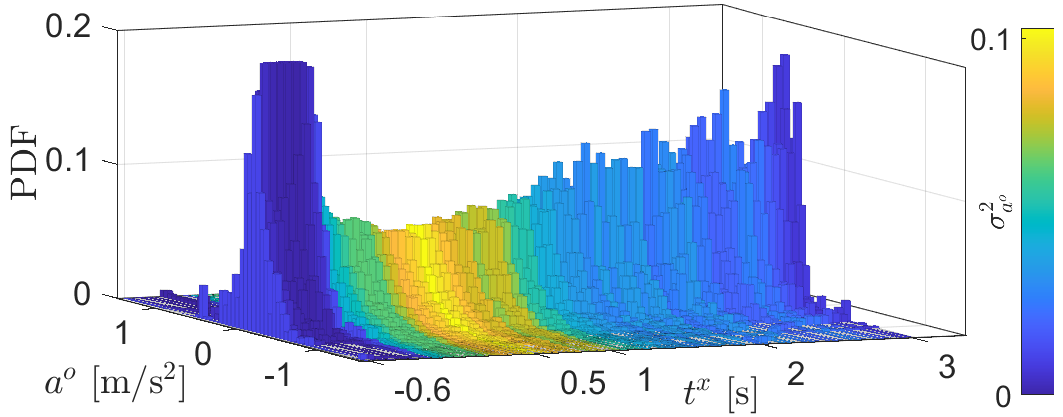}
         \label{fig:3D_gaussian}
     \end{subfigure}
     \begin{subfigure}{\columnwidth}
         \centering
         \includegraphics[width=\columnwidth]{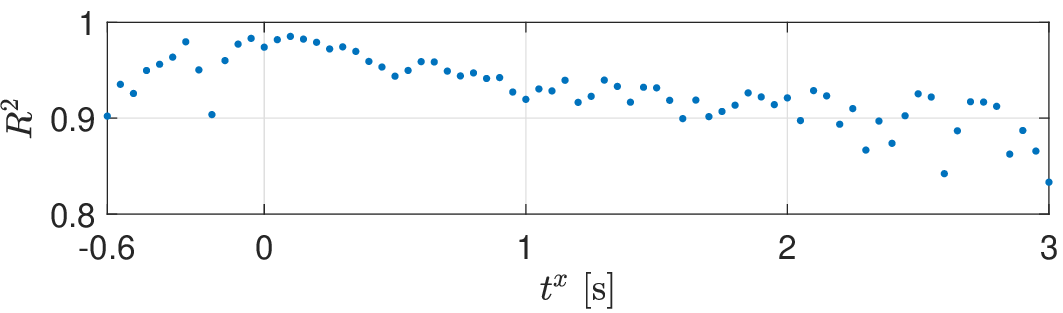}
         \label{fig:OV_gaussian_fitting}
     \end{subfigure}
\caption{Top plot: Three-dimensional probability density function (PDF) ($z$-axis) of HDV acceleration/deceleration $a^{o}$ ($y$-axis) for various values of relative longitudinal headway time $t^{x}$ ($x$-axis). 
Bottom plot: Coefficient of Determination ($R^2$) evaluating the accuracy of Gaussian fits to the acceleration/deceleration distributions across varying longitudinal headway times. 
}
\label{fig:OV_gaussian}
\end{figure}

\begin{figure}[t!]
\centering
\includegraphics[width=\columnwidth]{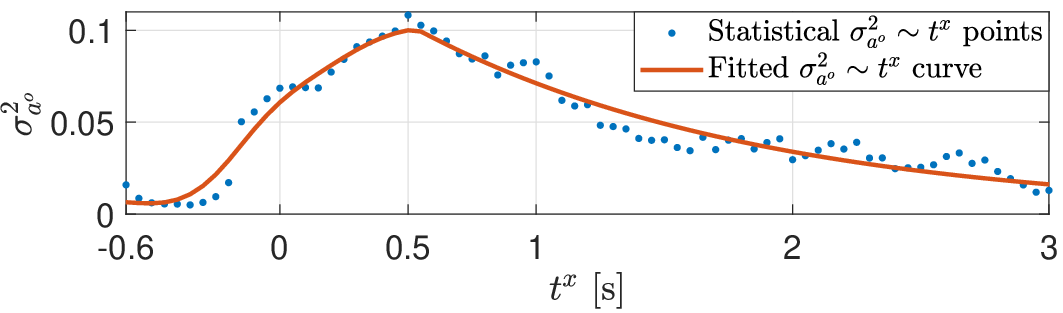}
\caption{Data points from statistical analysis and fitted curve of the variance of HDV acceleration/deceleration ($\sigma^{2}_{a^{o}}$) at different longitudinal headway times ($t^{x}$), with $R^2=0.9704$ and $R^2=0.9341$ for $-0.6\,\textrm{s}\leq t^{x}\leq 0.5\,\textrm{s}$ and $0.5\,\textrm{s}< t^{x}\leq 3\,\textrm{s}$, respectively.}
\label{fig:OV_gaussian_variance}
\end{figure}

To model the stochastic characteristic of the human-dominated uncertainty, the present work analyzes real-world HDV behaviors when they are overtaken based on the real-world highD dataset \cite{highDdataset} which records more than 110,000 naturalistic vehicle (car and truck) position trajectories at 6 different locations on German highways. 
Analysis of the position trajectories as well as velocity and acceleration profiles for each overtaking-overtaken pair indicates that overtaken vehicles tend to respond to the overtaking vehicle by adjusting their velocity when their longitudinal headway time ($t^{x}(t)=s^{x}(t)/v^{o}(t)$) falls within the range of $-0.6$\,s to $3$\,s, with a small negative longitudinal headway time signifying that the overtaking vehicle is still on the side and (just) behind the overtaken one. Meanwhile, for the OV longitudinal headway time that is out of this range, the overtaking vehicle has a negligible influence on the overtaken vehicle's behavior. 
Statistical analysis of the dataset within the range $-0.6\,\textrm{s}\leq t^{x}(t) \leq 3\,\textrm{s}$ reveals that the acceleration/deceleration distribution of the overtaken vehicle at each specific longitudinal headway time adheres to a Gaussian distribution profile, according to high coefficient of determination ($R^{2}$) results, especially for the period of $-0.6\,\textrm{s}\leq t^{x}(t) \leq 2\,\textrm{s}$ where $R^2$ is above 0.9, as illustrated by the three-dimensional probability density function (PDF) and $R^{2}$ plots in Fig. \ref{fig:OV_gaussian}.       
Moreover, the variance ($\sigma^{2}_{a^{o}}(t)$) of the Gaussian distribution varies with the longitudinal headway time ($t^{x}(t)$), as illustrated in Fig. \ref{fig:OV_gaussian_variance}.
The statistical analysis reveals that the variance increases monotonically to a peak when the longitudinal headway time ranges from $-0.6$\,s to $0.5$\,s, a period during which the overtaking vehicle passes alongside the overtaken vehicle.
The variance reaches its maximum at $0.5$\,s, suggesting that the uncertainty is highest at this point, which aligns well with the real-world driving experience, wherein the behavior of the overtaken vehicle becomes particularly unpredictable when the overtaking vehicle is marginally ahead of it. Subsequently, in the range $0.5\,\textrm{s}\leq t^{x}(t) \leq 3\,\textrm{s}$, the variance decreases exponentially as the overtaking vehicle progresses further ahead of the OV. 
The explicit numerical relationship between the variance and the longitudinal headway time is fitted accurately by a smoothing spline and an exponential function, for $-0.6\,\textrm{s}\leq t^{x}(t) < 0.5\,\textrm{s}$ and $0.5\,\textrm{s}\leq t^{x} \leq 3\,\textrm{s}$, respectively, with the fitted values of $R^2$ being 0.9704 and 0.9341, respectively, as plotted in Fig. \ref{fig:OV_gaussian_variance}.
For cases when $t^{x}(t)<-0.6\,\textrm{s}$ or $t^{x}(t)>3\,\textrm{s}$, $\sigma^{2}_{a^{o}}(t)$ is assumed to be extrapolated at the values corresponding to $t^{x}(t)=-0.6\,\textrm{s}$ and $t^{x}(t)=3\,\textrm{s}$, respectively, as longer longitudinal headway times have negligible effects according to the statistical data.
Compared with other stochastic approaches adopting theoretical distribution parameters of uncertainties 
\cite{Nguyen2017, Maki2022}, the chance constraint enforced by the proposed longitudinal control scheme, and that will be described next, is based on realistic statistical findings and a fitted numerical relationship with high accuracy, accounting for the uncertainty introduced by human behavior.  
\subsubsection{Formulation of the stochastic MPC}
According to the state evolution in \eqref{eq:lon_state_space_x}, since the Gaussian uncertainty can affect the $\hat{s}^{x}(k|t)$ state variable, instead of the normal collision avoidance constraint \eqref{eq:col_avod}, a chance constraint is formulated to take into account the stochastic uncertainties of the HDV behavior, as follows:
\begin{equation}\label{eq:lon_constraint_chances}
    \begin{aligned}
        &\text{Pr}\Bigl\{k^{x}(k|t)x^{x}(k|t)\leq b^{x}(k|t)\Bigr\}\geq 1-\beta,
    \end{aligned}
\end{equation}
where $\beta$ is a predefined risk coefficient. $k^{x}(k|t)=[\hat{k}_{p}(k|t),0,0]$ and $b^{x}(k|t)=s^{y*}(k|t)-\hat{b}_{p}(k|t)$, where $s^{y*}(k|t)$ is the optimal lateral position sequence generated by the lateral controller. $\hat{k}_{p}(k|t)$ and $\hat{b}_{p}(k|t)$ are determined analogously to the calculations of $\check{k}_{p}(k|t)$ and $\check{b}_{p}(k|t)$ in \eqref{eq:ov_OCP1}, with the only difference being that $v^{o*}(k|t)$ rather than $v^{o*}(k+1|t-1)$ is used when $p=3$. According to the property of the cumulative distribution function, given the mean ($a^{o*}(k|t)$) and variance ($\sigma^{2}_{a^{o}}(t)$) of the Gaussian uncertainty of the OV acceleration, the chance constraint \eqref{eq:lon_constraint_chances} can be converted into a deterministic form as 
\begin{equation} \label{eq:lon_constraint_chance_coll_avoid_yield}
    \begin{aligned}
        &\frac{1}{2}\left (1+\text{erf}\left(\frac{b^{x}(k|t)-\mu(k|t)}{\sqrt{2}\sigma(k|t)}\right)\right)\geq 1-\beta,
    \end{aligned}
\end{equation}
where $\mu(k|t)$ and $\sigma(k|t)$ are determined by the mean vector $\Pi(k|t)$ and covariance matrix $\Sigma(k|t)$ of the state $x^{x}(k|t)$ as
\begin{subequations} \label{eq:lon_constraint_chance_coll_avoid_vairance_and_mean}
    \begin{align}
        &\mu(k|t)=k^{x}(k|t)\Pi(k|t), \label{eq:lon_constraint_chance_coll_avoid_mean}\\
        &\sigma^{2}(k|t)=k^{x}(k|t)\Sigma(k|t)k^{x}(k|t)^{\top}, \label{eq:lon_constraint_chance_coll_avoid_vairance}
    \end{align}
\end{subequations}
where the initial $x^{x}(t)$ is deterministic, such that the mean vector of $x^{x}(t)$, $\Pi(t)=[s^{x}(t),v(t),v^{o}(t)]^{\top}$, is the actual system measurement, and the covariance matrix of $x^{x}(t)$ is $\Sigma(t)=\mathbf{0}^{3\times3}$. Due to the propagating property of the MPC algorithm, the dynamics of the mean and covariance of the state are, respectively,
\begin{subequations}\label{eq:lon_chance_constraint_propogating}
    \begin{align}
        \Pi(k+1|t)&=A^{x}\Pi(k|t)+B^{x}u^{x}(k|t)+E^{x}{a}^{o*}(k|t),\label{eq:lon_mean_propogating}\\
        \Sigma(k+1|t)&=A^{x}\Sigma(k|t)(A^{x})^{\top}+E^{x}\sigma^{2}_{a^{o}}(k|t)(E^{x})^{\top}. \label{eq:lon_sigma_propogating}
    \end{align}
\end{subequations}
Therefore, \eqref{eq:lon_constraint_chance_coll_avoid_yield} is equivalent to
\begin{subequations} \label{eq:lon_constraint_chance_coll_avoid_yield_equ}
    \begin{align}
        \left(b^{x}(k|t)-\mu(k|t)\right)^2 &\geq 2\sigma^{2}(k|t)\left(\text{erf}^{-1}(2\beta-1)\right)^2, \label{eq:lon_constraint_chance_coll_avoid_yield3}\\
        b^{x}(k|t)&\geq\mu(k|t). \label{eq:lon_constraint_chance_coll_avoid_yield4}
    \end{align}
\end{subequations}
In contrast to prior studies that conservatively assume that the human driver uncertainty is bounded \cite{Yu2024ECC}, this real-world data based statistical chance constraint approach 
mitigates excessive conservatism, thereby preserving a broader feasible solution space.
In addition to the chance constraint, the longitudinal controller is subject to associated state and control variable constraints, such that
\begin{subequations}\label{eq:lon_constraint_EV_va}
    \begin{align}
        \underline{v}&\leq \hat{v}(k|t)\leq \overline{v}, \label{eq:lon_constraint_EV_v}\\
        \underline{a}&\leq \hat{a}(k|t)\leq \overline{a}, \label{eq:lon_constraint_EV_a}
    \end{align}
\end{subequations}
where $\underline{v}$ and $\overline{v}$, and $\underline{a}$ and $\overline{a}$ denote EV longitudinal speed and acceleration limits, respectively.

The cost function of the longitudinal controller involves two objectives: driving the EV in a manner that gains a longitudinal distance advantage with respect to the OV while dynamically adapting to the real-time velocity of the OV to avoid unnecessary speeding, and minimizing the $L_{2}$-norm of the longitudinal acceleration magnitude to improve longitudinal comfort \cite{deWinkel2023} and minimize energy consumption \cite{Rabinowitz2024} during the overtaking maneuver. 
Hence, the objective function of the EV longitudinal controller is defined as 
\begin{equation} \label{eq:lon_objective_function_cost}
    \begin{aligned}
        \!\!J^{x}(t)\!=\!\sum_{k=0}^{N-1}\!-\hat{s}^{x}(k|t)P^{x}\!+\!\left(z^{x}\!(k|t)\!-\!\tilde{z}^{x}(k|t)\right)^{\!\top}\!Q^{x}\left(z^{x}\!(k|t)\!-\!\tilde{z}^{x}(k|t)\right),
    \end{aligned}
\end{equation}
where $P^{x} \in \mathbb{S}^{1}_{\succ}$, $Q^{x} \in \mathbb{S}^{2}_{\succ}$ 
determine the weighting of each cost in the multi-objective optimization. The cost vector $z^{x}(k|t)=[\hat{v}(k|t), \hat{u}^{x}(k|t)]^{\top}$ and the cost reference vector $\tilde{z}^{x}(k|t)=[v^{o}(t),0]^{\top}$. Due to the moving (relative space) coordinate system adopted in this work, the first (linear) term in the objective function directly motivates the EV to gain the distance advantage. 
Conversely, the speed target within the quadratic term prevents the EV from pursuing the distance target at its maximum velocity. However, once the OV accelerates to the point where the velocity difference between the two vehicles becomes negligible, the speed-target term in the cost function diminishes, re-emphasizing the distance-related term. This shift encourages the EV to re-establish a speed advantage, thereby extending its longitudinal lead over the OV. Consequently, the EV continuously optimizes its speed in real time while driving to the front of the OV. 
The acceleration target within the quadratic term of the objective function aims to minimize the $L_{2}$-norm of the acceleration throughout the entire task.
To summarize, by collecting the state-space equation in \eqref{eq:lon_state_space_x}, the chance constraint formulated in \eqref{eq:lon_constraint_chance_coll_avoid_yield_equ}, and inequality constraints in \eqref{eq:lon_constraint_EV_va}, the FHOCP addressing the longitudinal dynamics of EV is formulated as 
\begin{equation}\label{eq:lon_OCP1}
\begin{aligned}
\!\!\!\min\limits_{u^{x}(k|t)}\quad& J^{x}(t),\\
\!\!\!\textbf{s.t. } &\eqref{eq:col_avod_numerical_expressions},\eqref{eq:lon_state_space_x},\eqref{eq:lon_constraint_chance_coll_avoid_vairance_and_mean},\eqref{eq:lon_chance_constraint_propogating},\eqref{eq:lon_constraint_chance_coll_avoid_yield_equ},\eqref{eq:lon_constraint_EV_va}, \eqref{eq:lon_objective_function_cost}, \forall \; k\!\in\! \mathbb{N}_{[0, N-1]},\\
\!\!\!    \textbf{given: }& x^{x}(t)\!=\!\begin{bmatrix}
        s^{x}(t),v(t),v^{o}(t)
    \end{bmatrix}^{\top}, \sigma^{2}_{a^{o}}(k|t), a^{o*}(k|t),\\
    &v^{*}(k+1|t-1),v^{o*}(k|t), s^{y*}(k|t), \forall \; k\!\in\! \mathbb{N}_{[0, N-1]}.  
\end{aligned}
\end{equation} 
\begin{remark}\label{remark:3}
The present work assumes the EV speed limit is higher than that of OV ($\overline{v}>\overline{v}^{o}$) such that overtaking is feasible even when the OV aggressively speeds up. 
\end{remark}

\section{Simulation Results}
\label{sec:simulation_results}
\begin{figure}[t]
\centering
\includegraphics[width=\columnwidth]{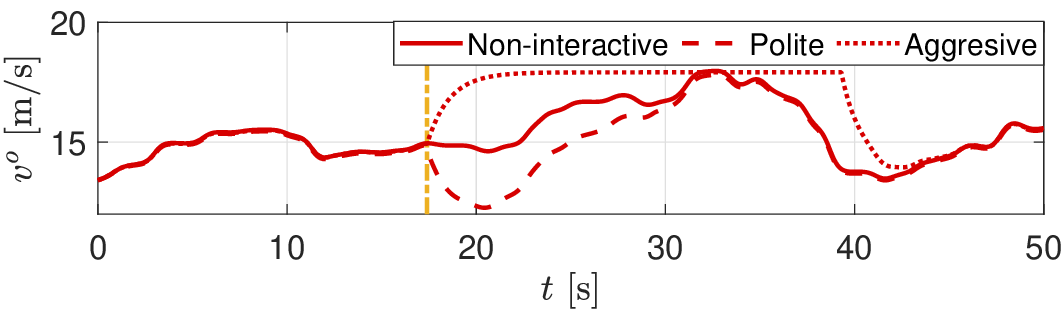}
\caption{The velocity profiles of non-interactive, polite, and aggressive OV. The interaction starts at 17.4\,s as indicated by the yellow dashed line.}
\label{fig:OV_velocities}
\end{figure}

The evaluation of the presented autonomous overtaking control scheme GT-PRO is carried out in three OV simulation scenarios: the EV overtaking interactive OVs with polite and aggressive responding behaviors, and a non-interactive OV, while comparing driving performances with state-of-the-art overtaking benchmark methods from the literature, given the same OV responses. In order to verify the robustness of the proposed control scheme, the simulations are based on the nonlinear vehicle dynamics in \eqref{eq:vehicle_bicycle_models}. In the non-interactive overtaking simulations, the OV follows a realistic speed profile, collected experimentally on a real road, and does not respond to the EV. 
Meanwhile, in the interactive overtaking simulations, the OV responds to the EV maneuver by adjusting its speed as the relative longitudinal distance between the two vehicles varies. To implement the interaction, an FHOCP similar to \eqref{eq:ov_OCP1} is adopted to mimic the OV longitudinal responses in simulations. 
While the weighting parameters of the GT-PRO in \eqref{eq:ov_OCP1} are tuned to represent the most common human driver responses, the weighting parameter values of the cost functions in the simulations are modified to model different human responses: by assigning either a smaller 
or a larger weighting parameter 
on the objective of tracking the desired speed profile, the solution of this optimization problem can represent polite or aggressive responses of human drivers, respectively, aligned with specific cases from the dataset.
Typical speed profile examples of a non-interactive, a polite, and an aggressive OV are illustrated in Fig. \ref{fig:OV_velocities}, respectively. 
As can be seen in that figure, the polite or aggressive OV driver brakes and accelerates, respectively, when the EV is marginally ahead of the OV, which is at approximately 17.4\,s in the examples shown in Fig. \ref{fig:OV_velocities}. In the polite scenario, the human driver, who prioritizes safety (i.e., the headway distance) as compared to tracking the speed target, intends to brake to yield space when the EV starts to merge in. In the case of the aggressive scenario, the human driver accelerates to the maximum speed when the EV starts to merge in, which challenges the EV's overtaking ability. Both the polite and aggressive drivers stop responding to the EV and continue to follow the initial speed profile when the EV is far ahead of the OV after the overtaking.
The EV speed limit ($\overline{v}$) is set higher than the OV speed limit ($\overline{v}^{o}$), set as the legal speed limit and the average traffic speed of the road, respectively, to ensure overtaking is theoretically feasible in this simulation. 

To comprehensively investigate the performance of the proposed GT-PRO method, the simulation section of this work consists of two parts. Section \ref{sec:simulation_results_1} demonstrates the OV and the controlled EV overtaking trajectories, including EV and OV velocities, and EV acceleration and steering angle, when overtaking a polite and an aggressive OV, respectively. 
Section \ref{sec:simulation_results_2} compares the driving performances between the proposed GT-PRO methods and the three following state-of-the-art benchmark methods from the literature, in polite and aggressive interactive, and non-interactive cases: 
\begin{itemize}
    \item Benchmark 1 method (BK1), which originates from the authors' previous work \cite{Yu2024ECC}, is a convex overtaking framework utilizing robust MPC and bounded uncertainty for the OV behavior, that can only manipulate the lateral motion of the EV, whereas longitudinal speed is fixed. 
    \item Benchmark 2 method (BK2) is adapted from \cite{Karlsson2020}, where a decoupled double-integral vehicle model is utilized to control the EV's longitudinal and lateral dynamics in overtaking. The approach is set up as a nonlinear control problem solved by mixed integer programming.
    \item Benchmark 3 method (BK3) is adapted from \cite{Gao2022}, which utilizes a nonlinear bicycle model to capture and manage both the longitudinal and lateral dynamics of the vehicle in overtaking.
    BK3 develops a risk-aware optimal overtaking algorithm based on martingale theory to address the uncertainty that human-driven OV can change its speed as a reaction to being overtaken. In addition, the reachability analysis-based safety conditions are reformulated into smooth and differentiable collision avoidance constraints.
\end{itemize}
\begin{table}[h!]
    \centering
        \caption{Parameters of the overtaking problem.}
    \label{tab:parameters}
    \begin{tabular*}{1\columnwidth}{c @{\extracolsep{\fill}} c@{\extracolsep{\fill}}c}
        \hline
        \hline
          Descriptions & Symbols & Values \\
         \hline
         EV/OV speed limit & $\overline{v}$/$\overline{v}^{o}$ & 19.67/17.88\,m/s \\
         Length/width/wheelbase of vehicle & $L_{v}$/$W_{v}$/$l$ & 4.4/1.82/2.5\,m\\
         Width of one lane & $W_{l}$ & 3.65\,m\\
         Standstill headway distance& $d_{x_{0}}$ & 6.08\,m\\
         Min/max accelerations of EV/OV & $\underline{a}$/$\overline{a}$ & -6.5/2.33 $\mathrm{m/s^2}$\\
         Min/max steering angle of EV & $\underline{\delta}$/$\overline{\delta}$ & $-5^{\circ}/5^{\circ}$ \\
         Min/max heading angle of EV & $\underline{\psi}$/$\overline{\psi}$ & $-5^{\circ}/5^{\circ}$\\
         Min/target longitudinal headway time& $\underline{t}$/$\tilde{t}$ & 1.5/2\,s\\
         Acceleration of Gravity& $g$ & 9.81\,$\mathrm{m/s^2}$\\
        \hline
        \hline
    \end{tabular*}
\end{table}
In the present simulations, the total simulation time is set to $T=50$\,s, with $N_T=500$ online iterations. The GT-PRO operates with a sampling interval of $\Delta T=0.1$\,s, and a prediction horizon length of $N=20$, resulting in a 2\,s prediction time, which provides a good balance between control performance and computational burden. 
The main characteristic parameters of the overtaking problem are summarized in Table \ref{tab:parameters}. 
The numerical simulations are carried out in the Matlab environment using the optimization toolbox Yalmip \cite{Lofberg2004} with MOSEK solver \cite{mosek} for BM1 methods, GUROBI solver \cite{gurobi} for the GT-PRO and BM2, and iPOPT solver \cite{ipopt} for BM3, respectively, on a 3.3 GHz Intel Core i7 processor with 16 GB onboard memory.
\begin{figure}[t!]
\centering
    \begin{subfigure}{\columnwidth}
         \centering
         \includegraphics[width=\columnwidth]{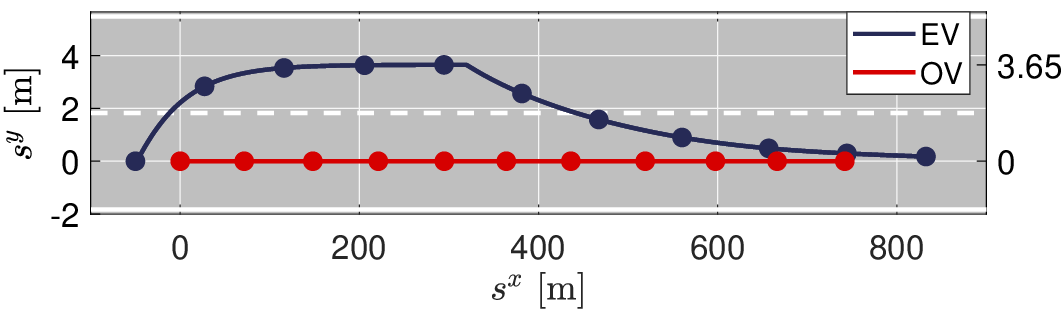}
    \end{subfigure}
    \begin{subfigure}{\columnwidth}
         \centering
         \includegraphics[width=\columnwidth]{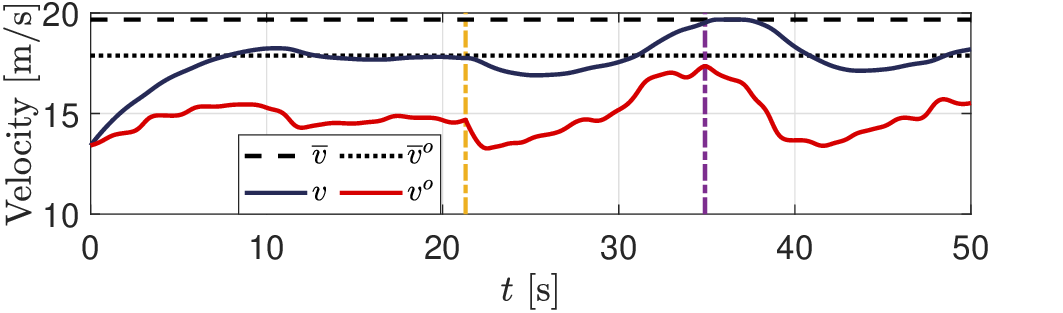}
     \end{subfigure}
     \begin{subfigure}{\columnwidth}
         \centering
         \includegraphics[width=\columnwidth]{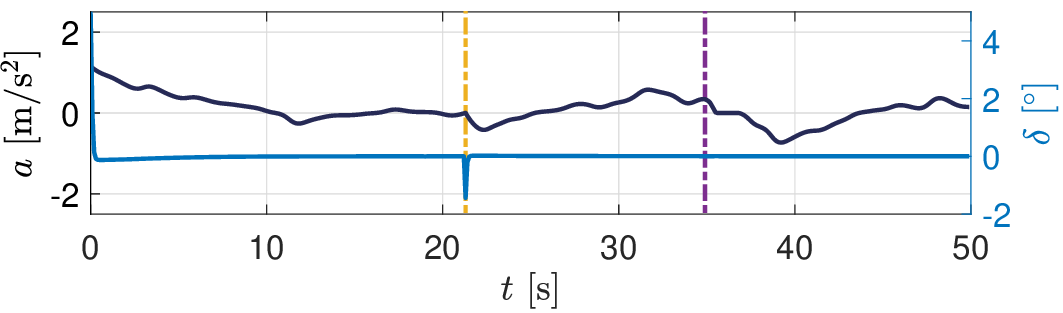}
     \end{subfigure}
\caption{Top to bottom: EV overtaking trajectory ($s^y$ against $s^x$ with each solid dot representing the vehicle's position at 5\,s time intervals),  EV and OV velocities ($v$, $v^{o}$), 
and EV acceleration and steering angle ($a$, $\delta$) for a polite OV case. Yellow and purple dashed lines indicate when $s^{x}(t)=s_{X_d}$, $s^{x}(t)=\tilde{s}^{x}(t)$, respectively, between which the OV reacts to the overtaking.}
\label{fig:Scenario_A_1}
\end{figure}

\subsection{EV and OV trajectories with GT-PRO method}
\label{sec:simulation_results_1}
In this section, unlike the common assumption in the literature where the EV initial speed is higher than that of the OV, the EV speed is initialized to be identical to the initial OV speed to comprehensively demonstrate not only the lateral but also the longitudinal performance of the GT-PRO method. 

\begin{figure}[t!]
\centering
    \begin{subfigure}{\columnwidth}
         \centering
         \includegraphics[width=\columnwidth]{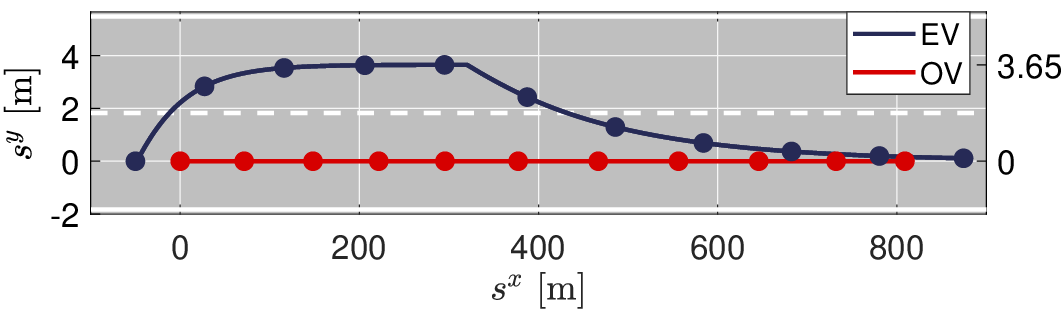}
    \end{subfigure}
    \begin{subfigure}{\columnwidth}
         \centering
         \includegraphics[width=\columnwidth]{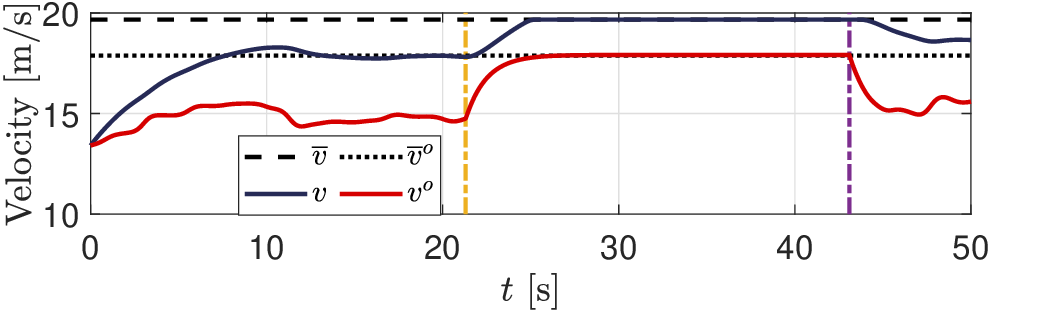}
    \end{subfigure}
    \begin{subfigure}{\columnwidth}
         \centering
         \includegraphics[width=\columnwidth]{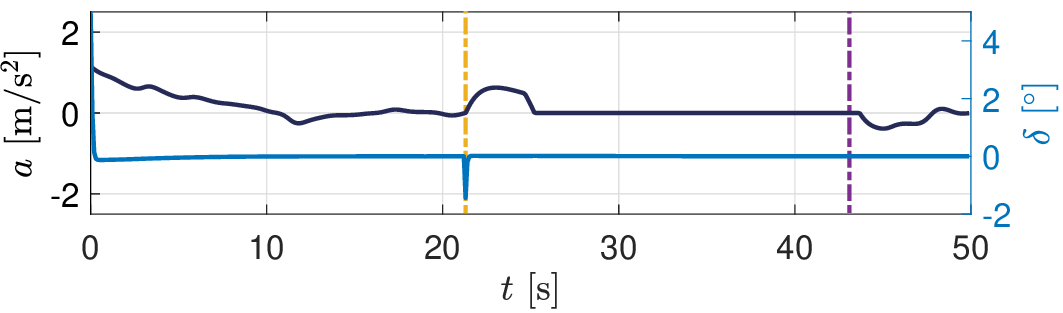}
    \end{subfigure}
\caption{Top to bottom: EV overtaking trajectory ($s^y$ against $s^x$ with each solid dot representing the vehicle's position at 5\,s time intervals),  EV and OV velocities ($v$, $v^{o}$), 
and EV acceleration and steering angle ($a$, $\delta$) for an aggressive OV case. Yellow and purple dashed lines indicate when $s^{x}(t)=s_{X_d}$, $s^{x}(t)=\tilde{s}^{x}(t)$, respectively, between which the OV reacts to the overtaking.}
\label{fig:Scenario_B_1}
\end{figure}

EV driving trajectories when overtaking a polite OV driver who yields space and an aggressive OV driver who speeds up while being overtaken are shown in Fig. \ref{fig:Scenario_A_1} and Fig. \ref{fig:Scenario_B_1}, respectively. 
The top subplot in Fig. \ref{fig:Scenario_A_1} depicts the trajectories of the EV and the OV, with 11 solid dots on each trajectory indicating the vehicles' instantaneous positions at each $T/10$ time interval, 
allowing for a direct comparison of their locations at corresponding times.
At the start of the overtaking scenario with the polite OV driver, the EV accelerates to and then maintains its speed around the traffic flow speed (i.e., $\overline{v}^{o}$) as illustrated in the middle subplot of Fig. \ref{fig:Scenario_A_1}. Due to the objective function design in \eqref{eq:lon_objective_function_cost}, the EV is able to actively adjust its speed to adapt to the varying OV velocity. For example, when the OV speeds up at around 32 s, the EV accelerates accordingly to ensure there is a sufficient speed difference between the two vehicles during the overtaking task.
The control variables of the lateral and longitudinal controllers are illustrated in the bottom subplot of Fig. \ref{fig:Scenario_A_1}. The EV steers to track the lateral position target and merges in at 21.3 s. On the other hand, the longitudinal controller tracks the speed target ($\overline{v}^{o}$), adaptively adjusting the velocity throughout the entire overtaking maneuver, irrespective of the overtaking phase, to guarantee safety.

Analogously, Fig. \ref{fig:Scenario_B_1} illustrates the EV overtaking trajectory when overtaking an aggressive OV that speeds up to prevent itself from being overtaken. As shown in the middle subplot of Fig. \ref{fig:Scenario_B_1}, when the OV aggressively speeds up to defend its position, the EV is able to accelerate further to the speed limit ($\overline{v}$) to maintain a sufficient speed difference such that the overtaking can be completed without collision, which illustrates the benefit of the adaptive longitudinal cost function design in \eqref{eq:lon_objective_function_cost}.
The control efforts generated by the longitudinal controller, which is adaptive to the real-time OV velocity, are shown in the bottom subplot of Fig. \ref{fig:Scenario_B_1}.

\subsection{Comparisons between GT-PRO and benchmark methods}
\label{sec:simulation_results_2}
\begin{figure}[t!]
\centering
    \begin{subfigure}{\columnwidth}
         \centering
         \includegraphics[width=\columnwidth]{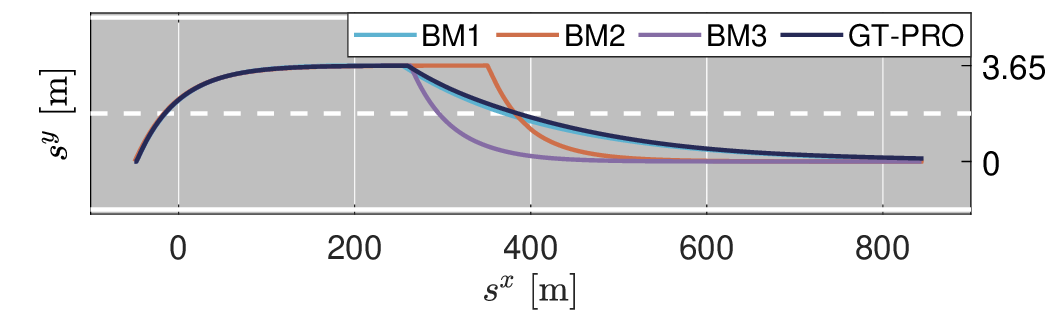}
    \end{subfigure}
    \begin{subfigure}{\columnwidth}
         \centering
         \includegraphics[width=\columnwidth]{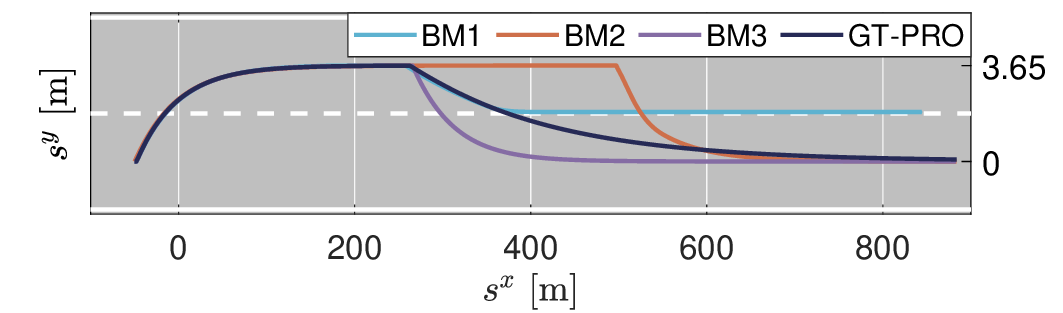}
    \end{subfigure}
    \begin{subfigure}{\columnwidth}
         \centering
         \includegraphics[width=\columnwidth]{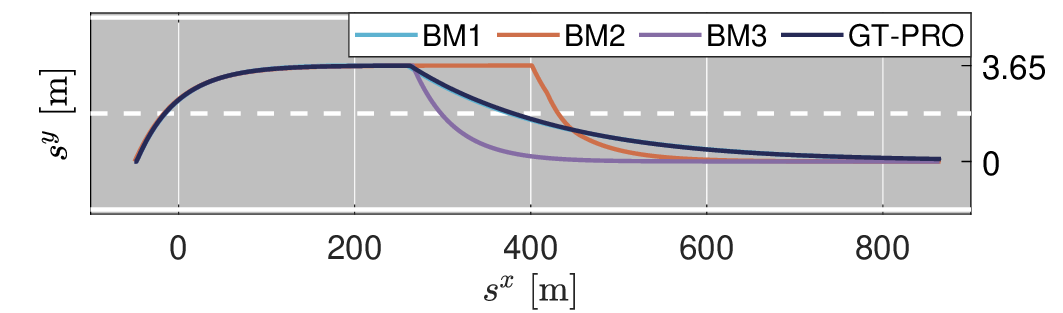}
    \end{subfigure}
\caption{Comparisons of the overtaking trajectories between benchmarks (BM1 \cite{Yu2024ECC}, BM2 \cite{Karlsson2020}, BM3 \cite{Gao2022}) and the GT-PRO when overtaking a polite OV (Case 1, top), an aggressive OV (Case 2, middle), and a non-interactive OV (Case 3, bottom). 
}
\label{fig:Scenario_2_023}
\end{figure}

In this section, we compare the overtaking performance between the proposed GT-PRO method and other autonomous overtaking methods existing in the literature: BK1 \cite{Yu2024ECC}, BK2 \cite{Karlsson2020}, and BK3 \cite{Gao2022} as introduced above. Here, the EV initial speed is initialized at the desired cruising speed so that the benchmark \cite{Yu2024ECC} that lacks the longitudinal speed control ability can be included as a baseline to demonstrate the improvement from the preliminary work. In addition to overtaking a polite OV (Case 1) and an aggressive OV (Case 2), this part further mimics a non-interactive OV (Case 3) that follows the pre-defined speed profile exactly to ensure the OV behaviors are identical for all methods in the comparison. OV velocity profiles of these three cases are shown earlier in Fig. \ref{fig:OV_velocities}. 
Moreover, to achieve a proper comparison, Case 3 is used to calibrate the weighting parameters of all methods (GT-RPO and benchmark methods), such that their pull-out trajectories, which are not affected by other surrounding vehicles, are as close as possible among methods, as illustrated in Fig. \ref{fig:Scenario_2_023}. Hence, the cut-in trajectories and associated driving performances of each method with different collision avoidance constraints and optimization designs can be fairly compared.    
Table \ref{tab:performance023} lists the root mean square (RMS) values of the heading angle ($\psi$) and lateral acceleration ($a^{y}$) performances of each method in the cut-in phase. 
The BM1 method exhibits a limitation when overtaking an OV that speeds up, e.g., in Case 2 as shown in the middle subplot of Fig. \ref{fig:Scenario_2_023}, where it fails to merge back to the middle of the initial lane. This shortcoming is caused by the lack of longitudinal control ability. Therefore, its driving performance in this specific case is not included in Table \ref{tab:performance023}. 
Numerical data show that the BM2 method results in the largest $\psi$ (0.908$^{\circ}$) and $a^{y}$ (0.711 $\mathrm{m/s^2}$) performances among all evaluated methods. 
The proposed GT-PRO method achieves the lowest RMS heading angle of 0.466$^{\circ}$ and lateral acceleration of 0.178 $\mathrm{m/s^2}$ on average, across all three OV behavior test cases. Compared with the BM1 method, the GT-PRO method achieves approximately 1\% and 2\% reductions on RMS $\psi$ and $a^{y}$, respectively. In addition, it is capable of completing overtaking even when OV speeds up to defend its position. Moreover, for the trajectory achieved by GT-PRO, the average RMS $\psi$ is reduced by 48.68\% and 48.05\% compared to BM2 and BM3 trajectories, respectively. Furthermore, the GT-PRO method yields 74.96\% and 71.75\% reductions on RMS $\delta$ with respect to BM2 and BM3, respectively. 
These findings underscore that the proposed GT-PRO method is demonstrated to produce more comfortable overtaking trajectories across various OV behavior cases, outperforming the benchmarks in the literature.

\begin{table}[h!]
    \centering
        \caption{Comparisons of the heading angle and lateral acceleration performances between methods.
        }
    \label{tab:performance023}
    \begin{tabular}{c|c|c|c|c|c}
\hline
\hline
 \multicolumn{2}{c|}{Methods}&BM1 \cite{Yu2024ECC}&BM2 \cite{Karlsson2020}& BM3 \cite{Gao2022}&GT-PRO \\ 
\hline
\multirow{2}{*}{Case 1}&$\psi$ [$^{\circ}$] & 0.469 & 0.938 & 0.904 & 0.467 \\ \cline{2-6}
& $a^{y}$ [$\mathrm{m/s^2}$] & 0.181 & 0.627 & 0.626 & 0.177 \\ 
\hline
\multirow{2}{*}{Case 2}&$\psi$ [$^{\circ}$] & - & 0.892 & 0.889 & 0.465 \\ \cline{2-6}
&$a^{y}$ [$\mathrm{m/s^2}$]  & - & 0.748  & 0.632 & 0.179 \\
\hline
\multirow{2}{*}{Case 3}&$\psi$ [$^{\circ}$] & 0.472 & 0.895 & 0.897 & 0.466 \\ \cline{2-6}
&$a^{y}$ [$\mathrm{m/s^2}$] & 0.183 & 0.758 & 0.633 &  0.179 \\ 
\hline
\multirow{2}{*}{Average}&$\psi$ [$^{\circ}$] & 0.471 & 0.908 & 0.897 & 0.466 \\ \cline{2-6}
&$a^{y}$ [$\mathrm{m/s^2}$] &0.182 & 0.711 & 0.630 &  0.178 \\ 
\hline
\hline
\end{tabular}
\end{table}
\begin{table}[h!]
    \centering
        \caption{Comparisons of the overtaking lane occupied time between methods.
        }
    \label{tab:occupied time}
    \begin{tabular}{c|c|c|c|c}
\hline
\hline
Methods& BM1 \cite{Yu2024ECC} & BM2 \cite{Karlsson2020} & BM3 \cite{Gao2022} & GT-PRO\\ 
\hline
Case 1& 21.6 s& 22.9 s& 17.5 s& 23.0 s \\ \hline
Case 2& 48.1 s& {29.2} s& 17.5 s& {21.4} s\\ \hline
Case 3& 21.9 s& {24.9} s& 17.6 s& {22.1} s\\
\hline
\hline
\end{tabular}
\end{table} 
\begin{table}[h!]
    \centering
        \caption{Comparisons of the minimum longitudinal headway time after the EV merges back to the initial lane between methods.
        }
    \label{tab:timegap}
    \begin{tabular}{c|c|c|c|c}
\hline
\hline
Methods& BM1 \cite{Yu2024ECC} & BM2 \cite{Karlsson2020} & BM3 \cite{Gao2022} & GT-PRO\\ 
\hline
Case 1& 2.45 s & 2.05 s & 1.47 s & 2.42 s \\ \hline
Case 2& - & 1.65 s & {0.50} s & {1.19} s \\ \hline
Case 3& 1.68 s & 1.84 s & {0.93} s & {1.98} s \\
\hline
\hline
\end{tabular}
\end{table}
In addition to comfort, one can also observe from Fig. \ref{fig:Scenario_2_023} that the BM2 method results in the longest occupation (except the unsuccessful attempt by BM1 in Case 2) of the overtaking lane, which is a consequence of its speed-independent ``critical zone'' collision avoidance constraint. While this configuration enhances safety during the cut-in maneuver by building up more longitudinal gaps, it decreases traffic passing efficiency due to unnecessary overtaking lane occupancy. The time spent on the overtaking lane by BM2 is 29.2 s and 24.9 s, in Case 2 and Case 3, respectively, which are 36.45\% and 12.67\% longer than that of the GT-PRO method (21.4 s for Case 2 and 22.1 s for Case 3), as listed in Table \ref{tab:occupied time}. 
In contrast, the BM3 method abruptly cuts off the OV, leaving insufficient space and time for the OV to react.
Although this solution is theoretically safe, it can become risky in realistic high-speed traffic scenarios with uncertainties in vehicle dynamics. According to Table \ref{tab:timegap} that records the minimum longitudinal headway time after the EV merges back to the initial lane (denoted as minimum headway time in the rest of this paper), the minimum headway time of BM3 is 0.93\,s in Case 3, and further drops to 0.50\,s in aggressive Case 2, which is below the critical boundary of 0.8\,s \cite{Makridis2020}. On the other hand, the GT-PRO method yields 1.19\,s and 1.98\,s minimum headway time in Case 2 and Case 3, respectively, both of which satisfy the threshold required to ensure sufficient reaction time for the OV \cite{Green2000,Kesting2008}.  

\begin{figure}[htb]
\centering
\includegraphics[width=\columnwidth]{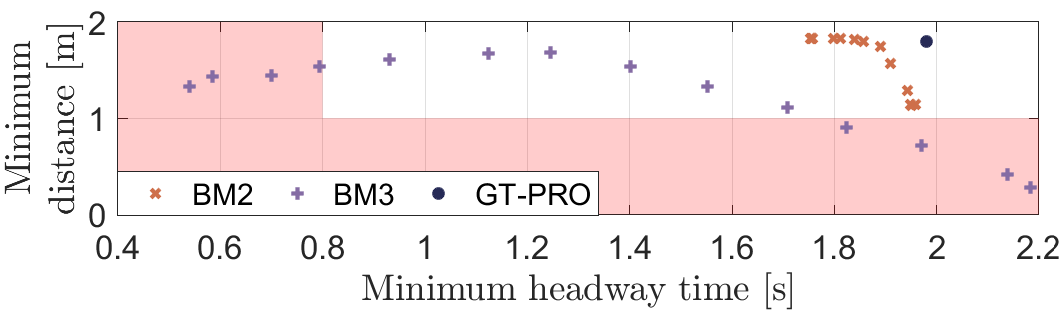}
\caption{Comparative Pareto analysis of minimum headway time and distance for the proposed GT-PRO method against benchmark BM2 and BM3 methods in Case 3, with the red-shaded region indicating the critical unsafe zone \cite{highDdataset,Makridis2020}.}
\label{fig:trajectory_tuning}
\end{figure}

In order to further elaborate on the safety-related performances of each method, a comparative Pareto analysis is carried out by varying the weighting parameters of optimizations in the benchmark methods. Fig. \ref{fig:trajectory_tuning} illustrates the Pareto analysis of the two benchmark methods, BM2 and BM3, in terms of the minimum time and minimum distance in overtaking, which is then compared with the proposed GT-PRO method. The red-shaded region denotes a critical unsafe zone with significant collision risks. This Pareto plot takes Case 3 as a representative example, where the simulated OV trajectory is identical for all methods. The results indicate that by varying weighting parameters, the benchmark methods (BM2 and BM3) can achieve performance levels comparable to the GT-PRO in either the minimum headway time or the minimum distance metric. However, this adjustment comes at the expense of the other metric, leading to a trade-off in overall performance. 
For example, the BM3 method can manage to yield a similar minimum headway time with the GT-PRO method at around 1.98\,s. On the other hand, the minimum distance between the two vehicles by BM3 with this specific tuning is around 0.72\,m, while it is at least 1.78\,m by the GT-PRO method. Based on the real-world dataset \cite{highDdataset} comprising more than 30,000 cutting-in trajectories, it is observed that the minimum lateral distance in the majority of safe cutting-in cases is greater than 1\,m.
In addition, Fig. \ref{fig:trajectory_tuning} shows that the BM3 method exhibits a wider spread due to its feasible region setup, with some instances in unsafe configurations. 
In contrast, the GT-PRO method maintains a balance between the minimum time and distance metrics, avoiding both excessive aggression and undue conservatism and staying outside the critical unsafe region, thus demonstrating a safer overtaking strategy. This comparison shows that while benchmark methods can be tuned to match GT-PRO in a specific safety metric, they often compromise the other safety metric.   

Regarding the computational effort, the GT-PRO method records an average computation time of 0.0579\,s per iteration, highlighting the potential for real-time implementation, while the nonlinear benchmark BM3 costs 1.4889\,s on average.   
\section{Conclusions}
\label{sec:conclusions}
In this work, the autonomous overtaking problem is addressed by the proposed Game-Theoretic, PRedictive Overtaking (GT-PRO) framework. The ego vehicle (EV) is managed by two coupled model predictive (MPC) controllers operating on the decoupled longitudinal and lateral dynamics, respectively, such that the important system dynamics are still captured effectively and the MPC controllers do not require time-consuming nonlinear nonconvex optimization calculations. Specifically, the lateral dynamics controller adopts a dynamic Stackelberg game-based MPC algorithm to interactively control the EV lateral motion and predict the uncertain overtaken vehicle (OV) reaction, thus leading to a bilevel optimization problem. The longitudinal dynamics controller is developed as a stochastic MPC that addresses deviations between the game-theoretic predicted and actual OV responses by robustly enforcing the collision avoidance constraint as a chance constraint to improve safety, where the variance of the OV behavioral uncertainty is obtained after analyzing a comprehensive real-world dataset that records human-driven vehicle accelerations when they are being overtaken by other vehicles.
The performance of the proposed GT-PRO method is assessed and compared with other autonomous overtaking schemes in the literature. Simulation results demonstrate that the proposed GT-PRO method is capable of overtaking not only a polite OV driver who yields but also an aggressive driver who speeds up when the EV is passing the OV. 
Moreover, when compared with benchmark methods from the literature, the proposed GT-PRO method achieves a more comfortable, efficient, and safe overtaking trajectory by comprehensively controlling both the longitudinal and lateral vehicle dynamics.     
Future work includes an extension of the current stochastic controller that is specifically designed for Gaussian-distributed uncertainties into a distributionally robust stochastic controller that is capable of handling uncertainties with unknown distributions. An explicit best response function of the OV (follower) can also be investigated to replace the numerical optimization-based follower subgame when formulating the game-theoretic formulation. 


\bibliographystyle{IEEEtran}
\bibliography{reference}

\begin{thebibliography}{10}
\providecommand{\url}[1]{#1}
\csname url@samestyle\endcsname
\providecommand{\newblock}{\relax}
\providecommand{\bibinfo}[2]{#2}
\providecommand{\BIBentrySTDinterwordspacing}{\spaceskip=0pt\relax}
\providecommand{\BIBentryALTinterwordstretchfactor}{4}
\providecommand{\BIBentryALTinterwordspacing}{\spaceskip=\fontdimen2\font plus
\BIBentryALTinterwordstretchfactor\fontdimen3\font minus \fontdimen4\font\relax}
\providecommand{\BIBforeignlanguage}[2]{{%
\expandafter\ifx\csname l@#1\endcsname\relax
\typeout{** WARNING: IEEEtran.bst: No hyphenation pattern has been}%
\typeout{** loaded for the language `#1'. Using the pattern for}%
\typeout{** the default language instead.}%
\else
\language=\csname l@#1\endcsname
\fi
#2}}
\providecommand{\BIBdecl}{\relax}
\BIBdecl

\bibitem{Yu2024TTE}
S.~Yu, X.~Pan, A.~Georgiou, B.~Chen, I.~M. Jaimoukha, and S.~A. Evangelou, ``A real-time robust ecological-adaptive cruise control strategy for battery electric vehicles,'' \emph{IEEE Transactions on Transportation Electrification}, vol.~10, no.~3, pp. 7389--7404, 2024.

\bibitem{Hamed2024}
H.~Rezaee, K.~Zhang, T.~Parisini, and M.~M. Polycarpou, ``Cooperative adaptive cruise control in the presence of communication and radar stochastic data loss,'' \emph{IEEE Transactions on Intelligent Transportation Systems}, vol.~25, no.~6, pp. 4964--4976, 2024.

\bibitem{PanXiao2023}
X.~Pan, B.~Chen, L.~Dai, S.~Timotheou, and S.~A. Evangelou, ``A hierarchical robust control strategy for decentralized signal-free intersection management,'' \emph{IEEE Transactions on Control Systems Technology}, vol.~31, no.~5, pp. 2011--2026, 2023.

\bibitem{Lian2023}
J.~Lian, W.~Ren, D.~Yang, L.~Li, and F.~Yu, ``Trajectory planning for autonomous valet parking in narrow environments with enhanced hybrid {A*} search and nonlinear optimization,'' \emph{IEEE Transactions on Intelligent Vehicles}, vol.~8, no.~6, pp. 3723--3734, 2023.

\bibitem{Gao2022}
Y.~Gao, F.~J. Jiang, L.~Xie, and K.~H. Johansson, ``Risk-aware optimal control for automated overtaking with safety guarantees,'' \emph{IEEE Transactions on Control Systems Technology}, vol.~30, no.~4, pp. 1460--1472, 2022.

\bibitem{Guanetti2018}
J.~Guanetti, Y.~Kim, and F.~Borrelli, ``Control of connected and automated vehicles: State of the art and future challenges,'' \emph{Annual Reviews in Control}, vol.~45, pp. 18--40, 2018.

\bibitem{Rahman2023}
M.~M. Rahman and J.-C. Thill, ``Impacts of connected and autonomous vehicles on urban transportation and environment: A comprehensive review,'' \emph{Sustainable Cities and Society}, vol.~96, p. 104649, 2023.

\bibitem{Ortega2024}
J.~Ortega, M.~Ortega, K.~Ismael, J.~Ortega, and S.~Moslem, ``Systematic review of overtaking maneuvers with autonomous vehicles,'' \emph{Transportation Engineering}, vol.~17, p. 100264, 2024.

\bibitem{Lodhi2023}
S.~S. Lodhi, N.~Kumar, and P.~K. Pandey, ``Autonomous vehicular overtaking maneuver: A survey and taxonomy,'' \emph{Vehicular Communications}, vol.~42, p. 100623, 2023.

\bibitem{Dixit2018}
S.~Dixit, S.~Fallah, U.~Montanaro, M.~Dianati, A.~Stevens, F.~Mccullough, and A.~Mouzakitis, ``Trajectory planning and tracking for autonomous overtaking: State-of-the-art and future prospects,'' \emph{Annual Reviews in Control}, vol.~45, pp. 76--86, 2018.

\bibitem{Liyiqun2024}
Y.~Li, Z.~Chen, T.~Wang, X.~Zeng, and Z.~Yin, ``Data-driven hierarchical model predictive control for automated overtaking maneuver via gaussian process regression,'' \emph{IEEE Transactions on Vehicular Technology}, pp. 1--15, 2024.

\bibitem{Dixit2020}
S.~Dixit, U.~Montanaro, M.~Dianati, D.~Oxtoby, T.~Mizutani, A.~Mouzakitis, and S.~Fallah, ``Trajectory planning for autonomous high-speed overtaking in structured environments using robust {MPC},'' \emph{IEEE Transactions on Intelligent Transportation Systems}, vol.~21, no.~6, pp. 2310--2323, 2020.

\bibitem{Yuan2024}
D.~Yuan, X.~Yu, S.~Li, and X.~Yin, ``Safe-by-construction autonomous vehicle overtaking using control barrier functions and model predictive control,'' \emph{International Journal of Systems Science}, vol.~55, no.~7, pp. 1283--1303, 2024.

\bibitem{Yuan2025}
S.~Yu, J.~Jiang, S.~Spurgeon, and B.~Chen, ``Safe control of autonomous vehicles in overtaking maneuvers using game-theoretic learning-based predictive controller,'' in \emph{2025 European Control Conference (ECC)}, 2025.

\bibitem{Gratzer2024}
A.~L. Gratzer, A.~Schmiedhofer, A.~Schirrer, and S.~Jakubek, ``Agile mixed-integer-based lane-change {MPC} for collision-free and efficient autonomous driving,'' \emph{IEEE Transactions on Intelligent Vehicles}, pp. 1--18, 2024.

\bibitem{Karlsson2020}
J.~Karlsson, N.~Murgovski, and J.~Sjöberg, ``Computationally efficient autonomous overtaking on highways,'' \emph{IEEE Transactions on Intelligent Transportation Systems}, vol.~21, no.~8, pp. 3169--3183, 2020.

\bibitem{Wang2024}
X.-F. Wang, W.-H. Chen, J.~Jiang, and Y.~Yan, ``High-level decision-making for autonomous overtaking: An {MPC}-based switching control approach,'' \emph{IET Intelligent Transport Systems}, vol.~18, no.~7, pp. 1259--1271, 2024.

\bibitem{Chai2023}
R.~Chai, A.~Tsourdos, S.~Chai, Y.~Xia, A.~Savvaris, and C.~L.~P. Chen, ``Multiphase overtaking maneuver planning for autonomous ground vehicles via a desensitized trajectory optimization approach,'' \emph{IEEE Transactions on Industrial Informatics}, vol.~19, no.~1, pp. 74--87, 2023.

\bibitem{Chen2020}
B.~Chen, D.~Sun, J.~Zhou, W.~Wong, and Z.~Ding, ``A future intelligent traffic system with mixed autonomous vehicles and human-driven vehicles,'' \emph{Information Sciences}, vol. 529, pp. 59--72, 2020.

\bibitem{Rahmati2022}
Y.~Rahmati, M.~K. Hosseini, and A.~Talebpour, ``Helping automated vehicles with left-turn maneuvers: A game theory-based decision framework for conflicting maneuvers at intersections,'' \emph{IEEE Transactions on Intelligent Transportation Systems}, vol.~23, no.~8, pp. 11\,877--11\,890, 2022.

\bibitem{Wei2022}
C.~Wei, Y.~He, H.~Tian, and Y.~Lv, ``Game theoretic merging behavior control for autonomous vehicle at highway on-ramp,'' \emph{IEEE Transactions on Intelligent Transportation Systems}, vol.~23, no.~11, pp. 21\,127--21\,136, 2022.

\bibitem{Lizhuoren2024}
Z.~Li, J.~Hu, B.~Leng, L.~Xiong, and Z.~Fu, ``An integrated of decision making and motion planning framework for enhanced oscillation-free capability,'' \emph{IEEE Transactions on Intelligent Transportation Systems}, vol.~25, no.~6, pp. 5718--5732, 2024.

\bibitem{Zhang2021}
X.~Zhang, A.~Liniger, and F.~Borrelli, ``Optimization-based collision avoidance,'' \emph{IEEE Transactions on Control Systems Technology}, vol.~29, no.~3, pp. 972--983, 2021.

\bibitem{Zhang2024}
Q.~Zhang, R.~Langari, H.~E. Tseng, S.~Mohan, S.~Szwabowski, and D.~Filev, ``Stackelberg differential lane change game based on {MPC} and inverse {MPC},'' \emph{IEEE Transactions on Intelligent Transportation Systems}, vol.~25, no.~8, pp. 8473--8485, 2024.

\bibitem{Stackelberg2011}
H.~von Stackelberg, \emph{\BIBforeignlanguage{eng}{Market Structure and Equilibrium}}, 1st~ed.\hskip 1em plus 0.5em minus 0.4em\relax Heidelberg: Springer Berlin, 2010.

\bibitem{Yu2024ECC}
S.~Yu, B.~Chen, I.~M. Jaimoukha, and S.~A. Evangelou, ``Game-theoretic model predictive control for safety-assured autonomous vehicle overtaking in mixed-autonomy environment,'' in \emph{2024 European Control Conference (ECC)}, 2024, pp. 3728--3733.

\bibitem{highDdataset}
R.~Krajewski, J.~Bock, L.~Kloeker, and L.~Eckstein, ``The high{D} dataset: A drone dataset of naturalistic vehicle trajectories on {G}erman highways for validation of highly automated driving systems,'' in \emph{2018 21st International Conference on Intelligent Transportation Systems (ITSC)}, 2018, pp. 2118--2125.

\bibitem{Kesting2007}
A.~Kesting, M.~Treiber, and D.~Helbing, ``General lane-changing model mobil for car-following models,'' \emph{Transportation Research Record}, vol. 1999, no.~1, pp. 86--94, 2007.

\bibitem{Rajamani2012}
R.~Rajamani, \emph{\BIBforeignlanguage{eng}{Vehicle dynamics and control}}, 2nd~ed., ser. Mechanical engineering series.\hskip 1em plus 0.5em minus 0.4em\relax New York: Springer, 2012.

\bibitem{Zheng2017}
Y.~Zheng, S.~E. Li, K.~Li, F.~Borrelli, and J.~K. Hedrick, ``Distributed model predictive control for heterogeneous vehicle platoons under unidirectional topologies,'' \emph{IEEE Transactions on Control Systems Technology}, vol.~25, no.~3, pp. 899--910, 2017.

\bibitem{Dunbar2006}
W.~B. Dunbar and R.~M. Murray, ``Distributed receding horizon control for multi-vehicle formation stabilization,'' \emph{Automatica}, vol.~42, no.~4, pp. 549--558, 2006.

\bibitem{deWinkel2023}
K.~N. {de Winkel}, T.~Irmak, R.~Happee, and B.~Shyrokau, ``Standards for passenger comfort in automated vehicles: Acceleration and jerk,'' \emph{Applied Ergonomics}, vol. 106, p. 103881, 2023.

\bibitem{Matute2019}
J.~A. Matute, M.~Marcano, S.~Diaz, and J.~Perez, ``Experimental validation of a kinematic bicycle model predictive control with lateral acceleration consideration,'' \emph{IFAC-PapersOnLine}, vol.~52, no.~8, pp. 289--294, 2019, 10th IFAC Symposium on Intelligent Autonomous Vehicles IAV 2019.

\bibitem{Luo1996}
Z.-Q. Luo, J.-S. Pang, and D.~Ralph, \emph{Mathematical Programs with Equilibrium Constraints}.\hskip 1em plus 0.5em minus 0.4em\relax Cambridge University Press, 1996.

\bibitem{Nguyen2017}
N.~A. Nguyen, D.~Moser, P.~Schrangl, L.~del Re, and S.~Jones, ``Autonomous overtaking using stochastic model predictive control,'' in \emph{2017 11th Asian Control Conference (ASCC)}, 2017, pp. 1005--1010.

\bibitem{Maki2022}
R.~Maki, I.~Okawa, and K.~Nonaka, ``Stochastic model predictive obstacle avoidance with velocity reduction in accordance with chance constraints in crowded environments,'' \emph{International Journal of Automotive Engineering}, vol.~13, no.~3, pp. 114--121, 2022.

\bibitem{Rabinowitz2024}
A.~I. Rabinowitz, C.~C. Ang, Y.~H. Mahmoud, F.~M. Araghi, R.~T. Meyer, I.~Kolmanovsky, Z.~D. Asher, and T.~H. Bradley, ``Real-time implementation comparison of urban eco-driving controls,'' \emph{IEEE Transactions on Control Systems Technology}, vol.~32, no.~1, pp. 143--157, 2024.

\bibitem{Lofberg2004}
J.~L{\"{o}}fberg, ``Yalmip : A toolbox for modeling and optimization in matlab,'' in \emph{2004 IEEE International Conference on Robotics and Automation (IEEE Cat. No.04CH37508)}, 2004, pp. 284--289.

\bibitem{mosek}
\BIBentryALTinterwordspacing
{MOSEK~ApS}, \emph{MOSEK Optimization Toolbox for MATLAB 10.0.43}, 2023. [Online]. Available: \url{https://docs.mosek.com/10.0/toolbox/index.html}
\BIBentrySTDinterwordspacing

\bibitem{gurobi}
\BIBentryALTinterwordspacing
{Gurobi Optimization, LLC}, ``{Gurobi Optimizer Reference Manual},'' 2024. [Online]. Available: \url{https://www.gurobi.com}
\BIBentrySTDinterwordspacing

\bibitem{ipopt}
A.~W\"{a}chter and L.~T. Biegler, ``On the implementation of an interior-point filter line-search algorithm for large-scale nonlinear programming,'' \emph{Mathematical Programming}, vol. 106, no.~1, pp. 25--57, 2006.

\bibitem{Makridis2020}
M.~Makridis, K.~Mattas, and B.~Ciuffo, ``Response time and time headway of an adaptive cruise control. {A}n empirical characterization and potential impacts on road capacity,'' \emph{IEEE Transactions on Intelligent Transportation Systems}, vol.~21, no.~4, pp. 1677--1686, 2020.

\bibitem{Green2000}
M.~Green, ```{H}ow long does it take to stop?' {M}ethodological analysis of driver perception-brake times,'' \emph{Transportation Human Factors}, vol.~2, no.~3, pp. 195--216, 2000.

\bibitem{Kesting2008}
A.~Kesting, M.~Treiber, M.~Sch\"{o}nhof, and D.~Helbing, ``Adaptive cruise control design for active congestion avoidance,'' \emph{Transportation Research Part C: Emerging Technologies}, vol.~16, no.~6, pp. 668--683, 2008.

\end{thebibliography}
\end{document}